\documentclass[12pt,preprint]{aastex}

\usepackage{CJK}

\usepackage{epsfig}

\shorttitle{Line Survey of AFGL\,2688}
\shortauthors{Zhang et al.}

\begin{document}

\title{A Molecular Line Survey of the Carbon-Rich Proto-Planetary Nebula AFGL\,2688 in the 3\,mm and 1.3\,mm Windows}

\begin{CJK*}{UTF8}{gbsn}

\author{Yong Zhang (张泳), Sun Kwok (郭新), Jun-ichi Nakashima (中岛淳一), Wayne Chau}

\affil{Department of Physics, University of Hong Kong, Pokfulam Road, Hong Kong, China}
\email{zhangy96@hku.hk; sunkwok@hku.hk; junichi@hku.hk}

\author{Dinh-V-Trung} 
\affil{Institute of Physics, Vietnam Academy of Science and Technology,  10 DaoTan Street, BaDinh, Hanoi, Vietnam}
\email{dvtrung@iop.vast.ac.vn}

\begin{abstract} 

We present a spectral line survey of the proto-planetary nebula \object{AFGL\,2688} in the frequency ranges of 71--111\,GHz, 157--160\,GHz, and 218--267\,GHz using the {\it Arizona Radio Observatory} 12\,m telescope and the {\it Heinrich Hertz Submillimeter Telescope}.  A total of 143 individual spectral features associated with  32 different molecular species and isotopologues were identified.  The molecules C$_3$H, CH$_3$CN, H$_2$CO, H$_2$CS, and HCO$^+$ were detected for the first time in this object.  By comparing the integrated line strengths of different transitions, we are able to determine  the rotation temperatures, column densities, and fractional abundances of the detected molecules. The C, O, and N isotopic ratios in AFGL\,2688 are compared with those in IRC+10216 and the Sun, and were found to be consistent with stellar nucleosynthesis theory.  Through comparisons of molecular line strengths in asymptotic giant branch stars, proto-planetary nebulae, and planetary nebulae, we discuss the evolution in  circumstellar chemistry in the late stages of evolution.

\end{abstract}


\maketitle
\end{CJK*}

\section{Introduction}

Circumstellar envelopes (CSEs) created by stellar winds from stars on the Asymptotic Giant Branch (AGB) stage are active sites of molecular synthesis. 
Thus far, more than 70 molecular species have been detected in CSEs \citep[][and references therein]{cer11}. During the post-AGB evolution,the effective temperatures of the central stars steeply increase, while at the same time, the remnants of the CSE expand and disperse.  The increasing ultraviolet (UV) radiation fields 
gradually dissociate the molecules, and finally ionize the envelopes, leading to the formation of planetary nebulae (PNs). The proto-planetary  nebula (PPN) is a rapid ($\sim10^3$ yr) transition phase  between the AGB and PN phases \citep{kwo93}. The chemical  processes during the evolution from the AGB to PN phases, however, are far from fully understood. Different carbon-to-oxygen abundance ratio sets up
different reaction routes. Although the molecular species in carbon-rich environments is generally thought to be more abundant than that in oxygen-rich environments, \citet{ziu07} recently found
unexpected complexity of the chemistry in oxygen-rich environments.   Unbiased spectral line survey is the best way to globally investigate the physical conditions and molecular composition in CSEs \citep[see][for a recent review]{cer11}.  Most of the work, however, has been  focused on the brightest carbon star IRC+10216  \citep[see, e.g.,][etc]{kaw95,cer00,cer10,pat11}. 
In order to better understand the processes involved in circumstellar chemistry, it would be useful to study objects at different stages of evolution, especially in the post-AGB and PN phases.


In order to understand the interactions between the molecular processes, stellar evolution, and chemical environment, our research group has undertaken a long-term project devoted to systematic molecular line surveys of a sample of evolved stars over a wide frequency range. We have reported the observations
of carbon-rich AGB stars IRC+10216, CIT\,6, AFGL\,3068, and a young PN NGC\,7027 in 1.3\,mm, 2\,mm, 3\,mm, and 7\,mm bands  \citep{he08,zy08,zy09a,zy09b,way12}. Since the same instrumental settings were used, the systematic uncertainties in the study of circumstellar chemistry are minimized.  Our results demonstrate that the evolved stars in the AGB stage have similar chemical compositions and the differences in the abundance of long carbon-chain and Si-bearing molecules can be explained by slightly different evolutionary status and carbon abundances.  In the PN stage, the chemical composition changes dramatically with the enhancement of ionized species and the destruction of long-chain molecules, implying the considerable influence of UV radiation field and/or shock waves on the circumstellar chemistry. The present paper is the sixth part of this series.  We present a 3\,mm and 1.3\,mm spectral
line survey toward the carbon-rich PPN AFGL\,2688, with the goal of investigating the chemical transition in the short evolutionary timescale  between the AGB and PN stages.

AFGL\,2688 (CRL 2688, V1610 Cyg, the Egg Nebula) is a bipolar PPN whose visible appearance consists of a pair of bipolar lobes, a dark equatorial lane, and a large number of roughly concentric arcs \citep{sah98b}.  The dark equatorial lane exhibits strong emission from H$_2$ and CO \citep{sah98a,cox00}, suggesting the presence of molecular matter.   Details of the stellar wind history and the nebular structures have been thoroughly discussed by \citet{bal12}, who obtained an expansion velocity of 15--18\,km\,s$^{-1}$.  The dynamical age of the nebula and mass loss rate were estimated to be $\sim$350 yr and $3\times10^{-5}$\,M$_\sun$\,yr$^{-1}$, respectively \citep{uet06,lo76}.   The central star of AFGL\,2688 is a completely obscured F-type supergiant with  an effective temperature of about 7250\,K and enhanced abundances of carbon, nitrogen, neon, and yttrium \citep{ish12}. The distance to this PPN is highly uncertain.  \citet{ney75} and \citet{coh77} estimated a distance of 1--1.5\,kpc,  while a smaller value (0.3--0.42\,kpc) was obtained by \citet{uet06} and
\citet{bal12} based on the expansion velocity. In lack of an accurate determination, we have adopted the commonly used value of 1\,kpc  \citep{sah98b}.

Because of their brightness in the infrared and strong molecular emissions, AFGL\,618 and AFGL\,2688 are two of the most well studied PPNs, and are therefore suitable for spectral line surveys at millimeter wavelengths. In terms of morphology, both objects are bipolar nebulae and \citet{wd05} suggested that the molecular richness observed in the two PPNs is due to the presence of dense and optically thick equatorial tori separating the bipolar lobes.  Based on the spectral type (B0) of the central star, AFGL\,618 is more evolved than AFGL\,2688 and is closer to becoming a PN. The line survey of AFGL\,618 has been performed by \citet{pardo07} with the IRAM 30\,m telescope in  the frequency range of 80--276\,GHz. For AFGL\,2688, most of previous observations were limited to specific molecular lines. \citet{park08} surveyed this object in the frequency range of 85--116\,GHz.  Recently, \citet{wes10} reported the {\it Herschel}-SPIRE FTS spectra of AFGL\,618 and AFGL\,2688 in the wavelength range from 195 to 670\,$\mu$m,
which trace the warm inner regions of the CSEs. Differing from AFGL\,618, AFGL\,2688 does not show emission from ionized species, suggesting that its cool central star has not yet provided enough UV photons to alter the chemistry of the CSE.

A comparison between the chemical compositions of the two PPNs can inform us  about the efficiency of chemical processes in PPN stage.  For this purpose, a spectral line survey of AFGL\,2688 over a wider frequency  range and with higher sensitivity is desirable.  As of 2013, the reported detected and tentatively detected  molecules in AFGL\,2688 include C$_2$, C$_3$, C$_4$, CO, CN, CS, C$_2$S, C$_2$H, C$_4$H, C$_5$H, C$_6$H, C$_4$H$_2$, C$_6$H$_2$, C$_3$N, HCN, HNC, HC$_3$N, HC$_5$N, HC$_7$N, HC$_9$N, HCP, PH$_3$, PN, NH$_3$, SiS, SiO, SiC$_2$, H$_2$, H$_2$O, NaCN, MgNC, AlF, and NaCl \citep[see][and references therein]{cer01,cer11,hig01,hig03,ten08,mil08,park08,ish12,wes10}.

The spectra of AFGL\,2688 presented in this paper represent the most complete line survey of this object to date. In Section~2, we describe the observations and the data reduction. In Section~3, we present the detected molecular lines, intensity measurements, rotation diagram analysis, calculations of 
molecular abundances, as well as isotopic ratios.  In Section~4, we discuss the implication of our observations on circumstellar  chemistry.  Finally, the conclusions are summarized in Section~5.

\section{Observations and data reduction}

The spectral line survey at the 3\,mm and 1.3\,mm windows were performed during the period from 2005 April and 2006 September with the {\it Arizona Radio Observatory (ARO)} 12\,m telescope at Kitt Peak and the {\it Heinrich Hertz Submillimeter Telescope (SMT)} 10\,m telescope at Mt. Graham.  The observations were conducted in beam switching mode, with an azimuth beam throw of {2\arcmin}. Pointing and focus were checked every two hours using nearby planets. We obtained the spectra of AFGL\,2688 in the frequency ranges of 71--111\,GHz, 157--160\,GHz\footnote{The 157--160\,GHz spectum was obtained using the 2\,mm receiver on the 12\,m telescope. Our original proposal is to also perform a complete 2\,mm line survey \citep[see][]{he08,zy09a,zy09b}. However, because of limited observation time and poor weather, only a narrow spectral range was observed at the 2\,mm window.}, and 218--267\,GHz, with an on-source  integration time of around one hour for each frequency setting. 
The dual-channel SIS receivers were employed in single sideband dual
polarization mode. The system temperatures were 150--400\,K and 400--700\,K for the 3\,mm and 1.3\,mm bands, respectively. The image rejection ratio was typically better than 20\,dB. The half-power beamwidths (HPBWs) of the {\it ARO} 12\,m telescope and the {\it SMT} are 86$''$--38$''$ and 32$''$--28$''$ over the frequency ranges 71--160\,GHz and  218--267\,GHz, respectively.  For the 3\,mm observations, the receiver back-ends were equipped with two 256-channel filter banks (FBs) with spectral resolutions of 500\,kHz and 1\,MHz and a  millimeter autocorrelator (MAC) with 3072 channels and 195\,kHz resolution. 
The 3\,mm data were calibrated to the $T^*_R$ temperature scale, which was corrected for atmospheric attenuation, radiative loss, and rearward and forward scattering and spillover by the chopper-wheel method. The main beam brightness temperature was derived using $T_{\rm R}=T^*_{\rm R}/\eta^*_m$, where $\eta^*_m$ is the corrected beam efficiency (0.94--0.71 over the frequency range 71--161\,GHz).
For the 1.3\,mm observations, the spectrometers used were a 2048-channel acousto-optical spectrometer (AOS) with a spectral resolution of 500\,kHz per channel and 1024-channel Forbes Filterbanks (FFBs) with a spectral resolution of 1\,MHz per channel. The temperature scale was given in term of $T^*_A$, which was corrected for atmospheric attenuation. The main beam temperature was derived using $T_{\rm R}=T^*_A/\eta_{mb}$, where the main beam efficiency, $\eta_{mb}$, is $\sim0.7$. The conversion factors are 32.5--29.1\,Jy\,K$^{-1}$ and 35\,Jy\,K$^{-1}$ for 3\,mm and 1.3\,mm observations, respectively.

The spectra were reduced using the CLASS software package in GILDAS, which is developed and distributed by the Observatorie de Grenoble and IRAM.  The scans due to bad atmospheric conditions and receiver instabilities were discarded, and the calibrated spectral data were co-added using the $rms$ noise of each spectrum as weights. Spectral baselines were subtracted by fitting the line-free spectral regions to the low-order polynomial.  The 3\,mm and 1.3\,mm spectra were smoothed and rebinned to a frequency resolution of 1\,MHz and 3\,MHz, respectively. This led to a typical $rms$ noise levels of $<8$\,mK in main beam temperature unit for both spectra.

\section{Results}

\begin{figure}
\includegraphics[scale=0.9]{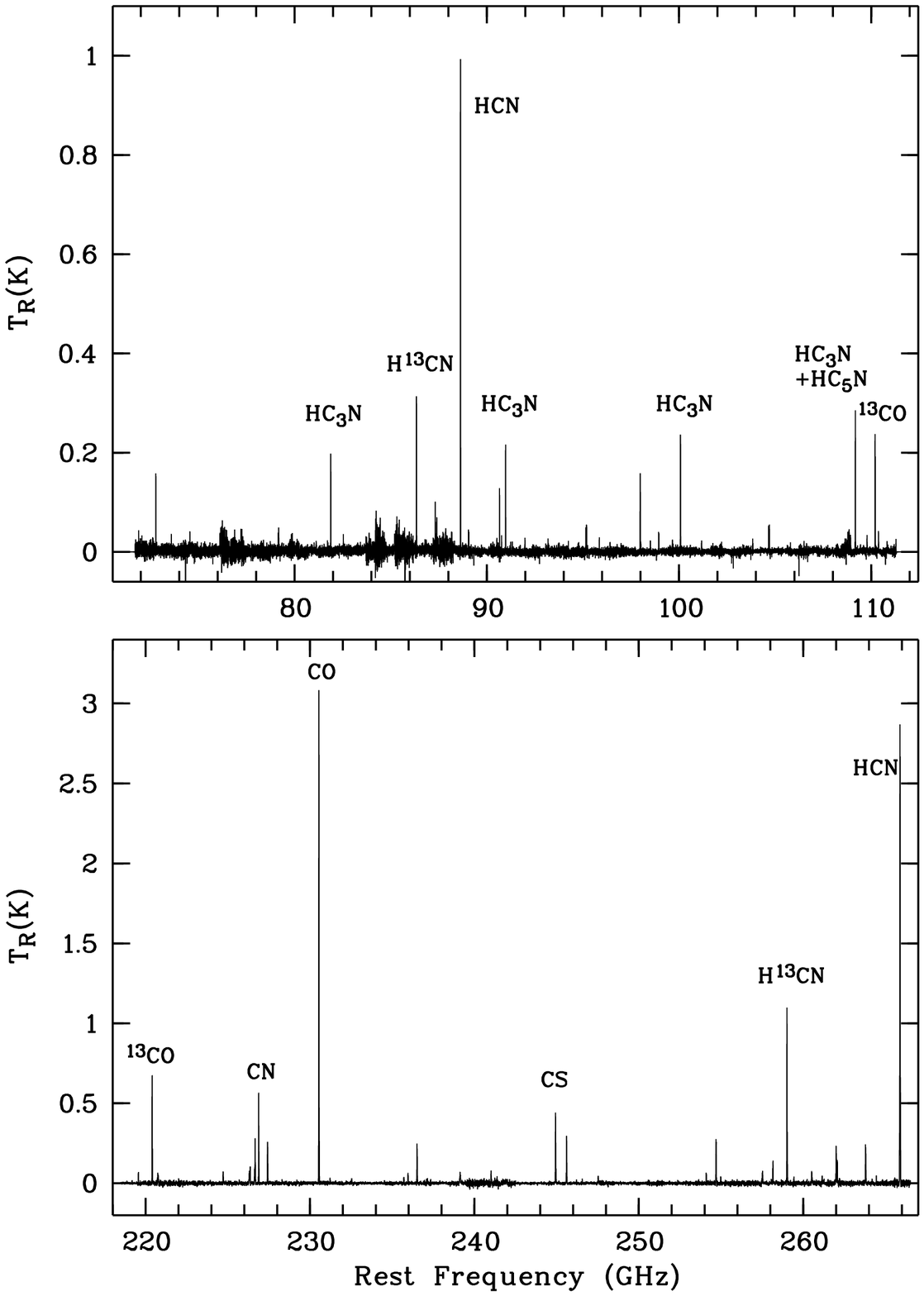}
\caption{Spectra covering the entire survey frequency range by the ARO 12 m (top) and the SMT (bottom).}
\label{overall}
\end{figure}

The 3\,mm and 1.3\,mm overall spectra are shown in Fig.~\ref{overall} with the strong lines identified.  More detailed spectra with expanded frequency scales for the 3\,mm and 1.3\,mm observations are shown in Fig.~\ref{spe_12m} and \ref{spe_smt}, respectively,  with identified lines marked in the spectra.  Complete data of the spectra are available in the on-line version of the Journal.  Figures~\ref{profile} shows the individual line profiles in velocity units.  Including uncertain detections, we detected a total of 143 lines (or groups of lines) belonging to 19 different molecular species and  13 isotopologues.  For the line detections, we have carefully examined the spectra to eliminate artifacts near the band edge.  
Although some lines are  faint (with intensities lower than 3$\sigma$ noise level), we still consider them as real or tentative detections because other transitions from the same species are clearly detected in this object or they have been commonly observed in other evolved stars. 
Most lines lie well within  $\pm20$\,km\,s$^{-1}$ of the central velocity. Unlike AFGL\,618,  no line is found to show P Cygni profiles.  Except SiS and SiO, all the observed molecules contain carbon. The line density is about 1.6 lines per GHz for both spectra.  Figure~\ref{cum} shows the cumulative number of detected lines exceeding a given main beam temperature. For comparison, we have also included the results from \citet{park08} in Figure~\ref{cum}.    We can see that many more lines are detected in the present work because of higher sensitivity and/or wider frequency range coverage.  The line identifications were made using the JPL \citep{pickeet98}, CDMS \citep{muller01,muller05}, and LOVAS \citep{lov92} spectral line databases.
To confirm our identifications, we checked the presence of other transitions of the same species at other frequencies, and compared the identifications with those from previous observations of other sources \citep{cer00,pardo07,he08}.  We can confidently identify all the features, except two lines at 233.938 GHz and 264.191 GHz.  Although these lines are close to 
the HC$^{13}$CCCCN ($J=$85--84) line at 223.937\,GHz and the Si$^{13}$CC (15$_{1,14}$--15$_{1,15}$) 
line at 264.190\,GHz, we do not make the association as the other transitions of HC$^{13}$CCCCN and Si$^{13}$CC are not detected. Table~\ref{crl} lists the line identification as well as the integrated line intensities ($\int T_{\rm R}$d$v$). The overall spectral properties are similar to that of IRC+10216.  No new species are detected in AFGL\,2688 beyond those seen in IRC+10216 and most of the non-detections could be due to the larger distance of AFGL 2688 compared to IRC+10216.


\subsection{Individual molecules}
\subsubsection{CO}

The $J=2$--1 transitions of CO, $^{13}$CO, C$^{17}$O, and C$^{18}$O as well as the $J=1$--0 transitions  of C$^{17}$O and C$^{18}$O were detected. Interferometric observations of the $^{13}$CO (1--0) line have revealed three components:  a bright central core with an expansion velocity of 19.6\,km\,s$^{-1}$, extensions with clumpy structure, and a high-velocity component \citep{yam95}.  The high angular resolution map of \citet{cox00}  shows that the CO (2--1) emission originates mostly from a central core with a diameter of $\sim$4\arcsec and multiple collimated outflows. As shown in Figure~\ref{profile}, the CO line has a parabolic shape, suggesting that it is optically thick.   There is a weak absorption feature at the blue wing ($\sim20$\,km\,s$^{-1}$), which is presumably caused by cold foreground gas in the optically dark lane at the center.  Showing rectangular or double-peak profiles, the isotopologue lines are less optically thick.   Using the {\it IRAM} telescope, \citet{bac90} obtained the main-beam brightness temperature of the CO (1--0) line to be about 15\,K at the center.  Considering the different beam sizes (12\arcsec\ for the {\it IRAM} observation), their measurement is consistent with ours. The integrated-intensity ratio of the CO and $^{13}$CO (2--1) lines is 4.07, in good agreement with that of the $J=1$--0 transitions (4.76) obtained by \citet{park08}.

\subsubsection{CS}

The $J=2$--1 and $J=5$--4 transitions of CS  were detected in the 2\,mm and 1.3\,mm windows, respectively. We also detected the isotopic transitions of $^{13}$CS, C$^{33}$S, and C$^{34}$S.  The C$^{33}$S emission is, however, extremely faint and its detection is only marginal. \citet{kas97} mapped the CS (2--1) line and found its emission peaking at both sides of the central dark lane, probably as the result of  enhancement of CS abundance by shock. They obtained a peak flux density of 4.5\,Jy, in good 
agreement with our observation (4.7\,Jy). Figure~\ref{profile} shows that both CS (2--1)  and (5--4) lines are optically thick.  The absorption feature observed in the blue wing of the CO (2--1) line is 
also present in the CS (2--1) line, but not seen in the higher $J$ transitions. We obtained the integrated intensity ratios of $I$(${^{12}}$CS 5--4)/$I$(${^{13}}$CS 5--4)$=18.3$ and  $I$(C${^{32}}$S 5--4)/$I$(C${^{34}}$S 5--4)$=6.74$, respectively, representing  the lower limits to the ${^{12}}$C/${^{13}}$C and ${^{32}}$S/${^{32}}$S abundance ratios.

\subsubsection{CN}

The cyanogen radical (CN) has one unpaired electron which splits the rotational states into doublets, each of which is split into hyperfine components because of the nuclear spin ($I=1$) of the N nucleus (Kwok 2007, section 7.10).   Three $N=2$--1 fine-structure groups ($J=3/2$--3/2, $J=3/2$--1/2, and $J=5/2$--3/2) of CN were clearly detected in the 1.3\,mm window. The strongest two groups have been detected by \citet{bac97a}.  We also observed two  $N=1$--0 groups ($J=1/2$--1/2, $J=3/2$--1/2) of $^{13}$CN
in the 3\,mm window. The $N=2$--1 groups are split into 18 hyperfine structure components, and can serve as a useful probe of the optical depth of molecular gas \citep{bac97b}.  Under the assumption of 
optically thin and local thermodynamic equilibrium (LTE) conditions, the intrinsic integrated intensity ratio between the three CN groups are $I_{\rm 226.8GHz}/I_{\rm 226.6GHz}=1.8$  and $I_{\rm 226.6GHz}/I_{\rm 226.3GHz}=5.0$, which are larger than the measured values of 1.4 and 1.7. Therefore, the two strongest CN groups are likely to have a large opacity. 

\subsubsection{C$_3$N and C$_n$H ($n=2$--$5$)}

We detected four pairs C$_3$N transitions and three pairs of C$_4$H transitions from $N=8$--7 to $N=11$--10 in the 3\,mm window. The C$_4$H (9--8) transition at 85.6\,GHz unfortunately falls within a spectral region with a high noise level. A vibrationally excited line of C$_4$H was marginally detected.
Each rotational transition of C$_3$N  and C$_4$H is split into two components with a similar intensity because of fine-structure interactions.   The C$_3$N (9--8) and C$_4$H (10--9) lines have been observed by \citet{luc86} and \citet{fukasaku94}. Considering different beam sizes, their intensity measurements are consistent with ours.   The spectra of C$_3$N  and C$_4$H show an optically thin and resolved
emission with an U-shaped profile (Figure~\ref{profile}). 

Five C$_4$H transitions from $N=24$--23  to $N=28$--27 fall within the 1.3\,mm window and are relatively strong in IRC+10216 \citep{he08}.  Although this molecular radical appears to be more abundant in AFGL\,2688 than in IRC+10216 \citep{luc86,fukasaku94}, none of these higher-$N$ transitions was detected in our spectra. This probably implies that the C$_4$H emission region of AFGL\,2688 is much colder than that of IRC+10216.  The first detection of ethynyl radical (C$_2$H) in    AFGL\,2688 was reported by  \citep{hug84}.  There are fine-structure groups ($N=1$--0 and $N=3$--2) of C$_2$H lying within the frequency range
of our line survey. Both of them were clearly detected.  The propynylidyne radical (C$_3$H) can be  linear ($l-$) or cyclic ($c-$) form.  Although faint, three blended features from the $^2\Pi_{3/2}$ state of $l-$C$_3$H were clearly detected in the 1.3\,mm window. $c-$C$_3$H is higher in energy and less stable than $l-$C$_3$H. The strongest $c-$C$_3$H features lying within the 3\,mm band are the 5(1,4)--4(1,3) transitions at 252.7\,GHz and 252.9\,GHz, which were not detected.  We also detected two faint features belonging to the $^2\Pi_{1/2}$ and $^2\Pi_{3/2}$ states of the pentadiynylidyne radical (C$_5$H) in the 1.3\,mm window.

\subsubsection{HNC}

The $J=$1--0  transition of HNC was clearly detected in the 3\,mm window.   The other HNC transitions lie outside the frequency range of our survey.   The $J=$1--0 and $J=$3--2 transitions of its isotopologue HN$^{13}$C fall within the spectral range.  We detected the $J=$3--2 transition in the 1.3\,mm window. The $J=$1--0 transition in the 3\,mm window is overwhelmed by noise.  \citet{park08}  suggested that the HNC (1--0) line is optically thick with a parabolic shape.  However, the present
more sensitive observations showed that the HNC emission is optically thin with a well-defined U-shaped profile (Figure~\ref{profile}).

\subsubsection{HC$_{2n+1}$N ($n=0$--$2$)}

The $J=$1--0 and $J=$3--2 transitions of HCN and its isotopologues H$^{13}$CN and HC$^{15}$N are strong lines in AFGL 2688.  \citet{bie88} mapped the HCN (1--0) line and found that the HCN emission is greatly enhanced in the optically dark lane and is not seen in the bipolar lobes, indicating a toroidal density distribution of this molecule. In our spectral survey, the HCN (3--2) and CO (2--1) lines have  comparable intensities and are the strongest lines in the spectrum. As shown in Figure~\ref{profile},
the profiles of HCN lines closely resemble to that of the CO line, showing an optically thick parabolic shape with an absorption feature at the blue wing. The profiles of isotopologue transitions appear more rectangular and are partially optically thin.  The integrated intensity ratio of H$^{12}$CN 
to H$^{13}$CN is $I$(H${^{12}}$CN 1--0)/$I$(H${^{13}}$CN 1--0)$=3.3$ and
$I$(H${^{12}}$CN 3--2)/$I$(H${^{13}}$CN 3--2)$=2.8$, suggesting that the higher-$J$ transition of H${^{13}}$CN is less optically thick.

Ten transitions in the ground vibrational state of HC$_3$N from $J=8$--7 to $J=29$--28 lie in the frequency range of our survey. All of them were prominently detected with $T_{\rm R}>150$\,mK. We also detected 12 vibrationally excited lines of HC$_3$N and 10 transitions of the isotopologues H$^{13}$CCCN, HC$^{13}$CCN, and HCC$^{13}$CN.  Figure~\ref{profile}  shows that these vibrationally excited lines have relatively narrow profiles, suggesting that they probably originate from inner warm regions.  To the best of our knowledge, only lower $J$ transitions of  HC$_3$N in the 3\,mm window have been detected in previous studies \citep[e.g.][]{fukasaku94,park08}.  The current detections of  HC$_3$N
transitions from levels covering a wider energy range allow us to more accurately derive physical conditions of the envelope.   

The integrated intensity ratios of HC$_3$N to H$^{13}$CCCN increase with increasing $J$ values, ranging from 11.8 for the $J=9$--8 transitions to 38.4 for the $J=26$--25. This is presumably ascribed to the effect of optical depth.

There are 15 HC$_5$N transitions from $J=28$--27 to $J=41$--40 in the frequency range of our survey. Except the $J=41$--40 transition which is badly blended with a HCN line, all of them were detected in
our spectra. Figure~\ref{profile} shows that the shape is more parabolic with the center part being quite flat, without the sharp edges typical of optically thin lines from expanding envelope.  The HCN ($J=$5-4) line observed with the VLA shows that the line is close to optically thick, with depression near systemic velocity due to the disruption of the HC3N hollow shell by the collimated outflows \citep{trung2009}.


\subsubsection{CH$_3$CN}

We report the first detection of  methyl cyanide (CH$_3$CN) in AFGL\,2688.  Previously,  the CH$_3$CN (1--0) transition has been searched for but only upper limits were reported  \citep{mat83}.    
Because of centrifugal distortion, each of the rotational transition $J+1 \to J$ of CH$_3$CN is split into $J$ different lines (Kwok 2007, section 7.8.2).  In the spectral range covered by our survey, there are a total of six groups consisting of 24 transitions from $J=4$--3 to $J=14$--13. All of them were evidently detected in spite of relatively weak in intensities.

\subsubsection{H$_2$CO and H$_2$CS}

We detected a H$_2$CO transition ($J_{K_{-1}, K_1}=3_{1,2}$--2$_{1,1}$) that is only marginally above the noise level. It is the strongest H$_2$CO line within the frequency range of our survey. We report a tentative detection of a H$_2$CS transition (7$_{1,6}$--6$_{1,5}$).  Both transitions were detected in the 1.3\,mm band with {\it SMT}, which provides a high sensitivity. This is the first detection of H$_2$CO and H$_2$CS in this object.

\subsubsection{HCO$^+$}

As far as we know, there is no previous report on the detection of ionized species in AFGL\,2688.
The $J=$1--0 transition of the molecular ion HCO$^+$  was previously searched for in AFGL\,2688, but only upper limits of 0.1\,K in main beam temperature were obtained  \citep{zuc76, bac97b}.  Here we present the first tentative detection of this species.  A weak peak at 89.2\,GHz is clearly visible in the 3\,mm window, and can be tentatively assigned to the HCO$^+$ $J=$1--0 line.

\subsubsection{SiS, SiO, and SiC$_2$}

In AFGL\,2688, lines from Si-bearing compounds are generally much fainter than those from C-bearing compounds. The SiS ($J=$5--4) transition was first detected in AFGL\,2688 by  \citet{fukasaku94}, but the line was not detected by \citet{park08}.  Four SiS transitions from $J=5$--4 to $J=14$--13 are in the frequency range of our survey.  All of them were detected with well-defined profiles.  We also detected the $J=2$--1 and $J=6$--5 transitions of SiO. The U-shaped profiles, as shown in Figure~\ref{profile}, indicate that these SiS and SiO lines are optically thin. 

SiC$_2$ is a triangular ring molecule and the SiC$_2$ ($J=$6--5) transition was detected in AFGL 2688 by  \citet{bac97b}. Our observations revealed a total of 21 SiC$_2$ transitions from $J=4$--3 to $J=12$--11. Emission lines from this refractory molecule are relatively faint, and most of them have less
well defined profiles, but the strongest SiC$_2$ line shows a double-peak   profile.

\subsection{Comparison with \citet{park08}}

The $\lambda3$\,mm spectral line survey of \citet{park08} resulted in the detections of eight molecules in AFGL\,2688. As illustrated in Figure~\ref{cum}, our spectra are about one order of magnitude deeper and are therefore able to reveal $\sim$5 times more lines than theirs. There is an overlapping frequency range (85--111\,GHz) between the survey of \citet{park08} and ours. In Figure~\ref{park}, we compare the integrated 
intensities of the lines detected in the overlapping region.  Because the beam sizes of the two observations are similar, we do not correct for the effect of beam dilution. Inspection of Figure~\ref{park} shows that the intensities of the lines detected in the two observations are in excellent agreement. Our observations detected all the species in those of \citet{park08}, except C$_2$S (shown in red).  There are 14 molecules detected in our survey (shown in blue) that were not detected by \citet{park08}.  

\citet{park08} claimed the first detection of C$_2$S in AFGL\,2688 through one transition ($J=8$--7). The intensity of this C$_2$S  line given by \citet{park08} is well above our detection limit. 
However, we did not detect this line with a 3$\sigma$ upper limit of 0.01\,K.   Furthermore, \citet{park08} did not detect the $J=7$--6 and $J=9$--8 transitions of C$_2$S which also lie in the frequency range of their survey and have  comparable strengths to the $J=8$--7 line in IRC+10216 \citep{cer87}.
Therefore, we suspect that this detection is not real, and C$_2$S cannot be detected in this object at the present sensitivity level.

\subsection{Rotation temperatures, column densities, and fractional abundances}

We have applied the rotation-diagram technique to derive the excitation temperatures ($T_{\rm ex}$) and column densities ($N$) of the molecules detected in our spectra, assuming LTE, optically thin condition and negligible background temperature.   In Figure~\ref{dia}, we have plotted the populations of the upper levels ($N_u$) against the corresponding excitation energies ($E_u$) in the transitions, using the equation,

\begin{equation}
\ln \frac{N_u}{g_u}=\ln\frac{3k\int T_s dv}{8\pi^3\nu S\mu^2}=
\ln\frac{N}{Q(T_{\rm ex})}-\frac{E_u}{kT_{\rm ex}},
\end{equation}
where $g_u$ is the degeneracy of the upper level, $S$ the line strength, $\mu$ the dipole moment, $\nu$  the line frequency, $Q$ the rotational partition function, and $k$ the Boltzmann constant. Assuming that the surface brightness has a Gaussian distribution, the source brightness temperature $T_s$  is given by $T_s=T_{\rm R}(\theta^2_b+\theta^2_s)/\theta^2_s$, where $\theta_b$ is the antenna HPBW and $\theta_s$ is source size. 
The $\theta_s$ value may be different for different molecules, and is poorly known for most of the species.  We followed the assumption by \citet{park08} as well as \citet{fukasaku94} and took an uniform 
$\theta_s$ of $20''$ for all the species. This assumption leads to a factor of two uncertainty in $T_s$.  If a given molecule has more than one transition detected, its $T_{\rm ex}$ and $N$ can be deduced using a straight-line fit to ${N_u}/{g_u}$ versus ${E_u}/{kT_{\rm ex}}$. When more transitions are detected, the results are more reliable. Departure from the linear relation may imply non-LTE excitation, different excitation mechanisms, effect of optical thickness, misidentification, and/or line blending.  The effects of saturation and subthermal excitation on the rotation diagram
have been extensively discussed by \citet{gol99}.
Since our present survey covers a wide frequency range and our detections include molecular transitions arising from energy levels over a wide energy range, the rotational diagram analysis can be particularly useful.

The rotation diagrams for ten molecules detected in our survey are shown in Figure~\ref{dia}.  The derived values of  $T_{\rm ex}$ and $N$ are given in Table~\ref{col_crl2688}, together with the results by \citet{fukasaku94} and those of IRC+10216 \citep{he08} for comparison. $T_{\rm ex}$ may be different from species to species. Generally, there are good linear relationships between ${N_u}/{g_u}$ and ${E_u}/{kT_{\rm ex}}$. Nevertheless, some molecules (i.e. SiC$_2$, HC$_3$N, and CH$_3$CN) appear to arise from regions with temperature variations. The average $T_{\rm ex}$ of HC$_3$N is 46.1\,K, higher than the 25.8\,K value obtained by \citet{park08} using only $\lambda3$\,mm data.  Figure~\ref{dia} clearly shows that the higher-$J$ transitions of HC$_3$N have a higher $T_{\rm ex}$. HC$_5$N transitions were detected only in the 3\,mm band and have a low $T_{\rm ex}$ value of 27.8\,K, in good agreement with the that of HC$_3$N deduced by \citet{park08}.  This implies that the cyanopolyynes coexist in the same regions.   The deduced column densities are in order-of-magnitude agreement with the results of \citet{fukasaku94}.  \citet{gol99} suggested that finite opitical depth $\tau$ can cause  deviations of the points from a straight line  on the rotation diagram, and an optical correction factor $C_\tau=\tau/(1-e^{-\tau})$ should be introduced in the rotation diagram. However, using the ``population diagram'' technique presented by \citet{gol99}, we found that the $\ln C_\tau$ values for all the species in Figure~\ref{dia} are too small ($<0.02$) to affect our results.

We also calculated the fractional abundances ($f_X$)  of the detected molecules relative to H$_2$ using the formula proposed by \citet{olo96}. A detailed explanation of the method can be found in \citet{zy09b}.  The deduced $f_X$ values and those normalized to HC$_3$N are given by Table~\ref{col_crl2688}, where the results for IRC+10216 and AFGL\,618 \citep{woods03,pc07} are listed for comparison.  It should bear in mind that the calculations were based on assumed mass-loss rate, distance, and
geometry of the shell, which are not well known.  We estimate that the
error of $f_X$ amounts to a factor of 5. Nevertheless, when comparing
the relative $f_X$ between detected molecules, the uncertainties invoked
by these assumed parameters can be greatly reduced.

\subsection{Isotopic ratios}

The circumstellar envelopes of evolved stars are constantly replenished by products of nucleosynthesis in the stellar interior. As a star evolves, changes of nuclear process changes lead to changing isotopic compositions, which are eventually reflected in the composition of the CSE.
The isotopic ratios in the CSE of IRC+10216 have been obtained by several groups \citep[i.e.][]{wan91,kah00}. These studies suggest that the $^{12}$C/$^{13}$C,  $^{16}$O/$^{17}$O, and  $^{18}$O/$^{17}$O ratios are lower compared to the solar values, while the $^{14}$N/$^{15}$N ratio is more than 10 times larger than solar.  The $^{16}$O/$^{18}$O ratio and those of heavier elements are consistent with the solar values.
We have determined the isotopic ratios of C, N, O, and S in AFGL\,2688 through the same $J$ transitions of molecules and their isotopologues. The calculations are based on the assumptions that the main and isotopic lines are  optically thin  and the lines are  under  the same excitation conditions.  The results, along with those of IRC+10216 and the Sun, are presented in Table~\ref{isot}.  The main lines of CO, HCN, and CS are optically thick, and thus only give the lower limits of the corresponding isotopic ratios.

We have detected the rare species $^{12}$C$^{34}$S and $^{13}$C$^{32}$S. The $^{12}$C$^{34}$S/$^{13}$C$^{32}$S ratio is $2.7\pm1.3$. Assuming the that $^{32}$S/$^{34}$S ratio is solar, we obtained $^{12}$C/$^{13}$C$=61\pm30$. $^{12}$C/$^{13}$C can also be obtained through the HC$_3$N and isotopic lines which are likely to be optically thin.  A comparison the two results suggests a $^{12}$C/$^{13}$C ratio of $\sim$31--37, slightly lower than that of IRC+10216 and significantly lower than the solar value. We note that the results obtained by previous authors scatter over a wide range of 20--66 \citep[see][]{mil09}.   The low $^{12}$C/$^{13}$C ratio is  typical for carbon-rich CSEs and is suggestive of the existence of a nonstandard mixing mechanism \citep[or cool bottom processing; e.g.,][]{charbonnel95} during the red giant phase of low-mass AGB stars.  A more detailed discussion about the implications of $^{12}$C/$^{13}$C  for AGB nucleosynthesis can be found in \citet{mil09}.

Although not strictly true, the derived $^{16}$O/$^{17}$O abundance ratio is expected to be positively correlated with that of $^{12}$C/$^{13}$C since both $^{13}$C and $^{17}$O are produced by $^{12}$C and $^{16}$O through ($p,\gamma$) reactions followed by $\beta$ decay.  Assuming that the $^{13}$CO and C$^{16}$O lines are optically thin and the $^{12}$C/$^{13}$C ratio is smaller than the solar value, we derive a $^{16}$O/$^{17}$O ratio of $<$810 in AFGL\,2688, or at least three times smaller than the solar value.

From the $J=2$--1 transitions of C$^{18}$O and C$^{17}$O, we obtain an $^{18}$O/$^{17}$O ratio of
$0.8\pm0.2$, a value consistent with that in IRC+10216 and significantly lower than the solar value, suggesting that $^{17}$O in this object has been substantially enhanced. Combining the present data with
those of other CSEs previously reported by us \citep{he08,zy09a,zy09b},  we can clearly see a positive correlation  between $^{16}$O/$^{17}$O and $^{12}$C/$^{13}$C ratios (Figure~\ref{isoto}).

On the other hand, $^{13}$C and $^{17}$O can be destroyed by the reactions $^{13}$C($p,\gamma$)$^{14}$N and $^{17}$O($p,\alpha$)$^{14}$N. Both paths lead to the overproduction of $^{14}$N. The enhancement of $^{14}$N in carbon-rich CSEs has been confirmed by the observations of \citet{wan91},  who obtained a
lower limit of 544 for the $^{14}$N/$^{15}$N ratio in AFGL\,2688, about twice higher than the solar value.  Assuming that the $J=$3--2 transitions of H$^{12}$C$^{14}$N and H$^{13}$C$^{15}$N  are optically
thin and the $^{12}$C/$^{13}$C is smaller than the solar value, we obtained an upper limit for $^{14}$N/$^{15}$N of 751, about 6 times lower than the lower limit in IRC+10216. The lower $^{14}$N/$^{15}$N value in AFGL\,2688 may be related to the slightly lower $^{12}$C/$^{13}$C  and $^{16}$O/$^{17}$O ratios than those in IRC+10216 (Figure~\ref{isoto}), suggesting that the two objects probably have different intrinsic properties, such as metallicity and mass.

The nucleosynthesis of sulfur isotopes $^{33}$S and $^{34}$S requires relatively high temperatures, which are are unlikely to be achieved in the interior of intermediate- and low- mass stars.  This is supported by our results that no discrepancy between the S isotopic ratios in AFGL\,2688 and the solar values is found.


\section{Discussion}

In order to investigate the chemical evolution of circumstellar envelopes during the AGB--PPN--PN transitions,  we have plotted the integrated-intensity ratios of the molecular lines in AFGL\,2688, IRC+10216 and NGC\,7027 in Figures~\ref{withirc} and \ref{withngc}, respectively. The $I_{\rm int}$(AFGL2688)/$I_{\rm int}$(IRC+10216) and $I_{\rm int}$(NGC7027)/$I_{\rm int}$(AFGL2688) ratios of the lines from the molecular 
species X [respectively represented as $R_1$(X) and $R_2$(X) hereafter] can, to some extent, reflect the relative abundance of X in these objects.  Although line intensity ratios can be affected by factors in addition to molecular abundance, such as different excitation conditions and optical depths, these effects can be substantially reduced by comparing lines arising from different levels. Therefore, our observations covering a wide frequency range are particularly suited for such a study.  Our results show that $R_1$(X) ranges from 0.02 to 2.89 with a mean value of 0.41, while $R_2$(X) ranges from 0.02 to 20.2 with a mean value of 1.64. Generally, the chemical pattern of AFGL\,2688 is more similar to that of IRC+10216 compared
to NGC\,7027, suggesting that the radiation field from the central star plays an important role on circumstellar chemistry and can alter chemical composition over a short time scale.  

IRC+10216 is the closest carbon-rich star \citep[$d\sim$120\,pc,][]{lou93}.  NGC\,7027 has a similar distance \citep[$\sim$880\,pc,][]{mas89} as AFGL\,2688.  If the three objects had nearly identical intrinsic brightness, 
the line intensities should be proportional to $(1+\theta_b^2/\theta_s^2)^{-1}d^{-2}$, as denoted by the dashed lines in Figures~\ref{withirc} and \ref{withngc}.  We can infer from Figures~\ref{withirc} and \ref{withngc} that the intrinsic molecular line strengths in AFGL\,2688 and NGC\,7027 are consistently
at least one order of magnitude higher than those in IRC+10216 if they are at the same distance from the earth. 
This suggests that the detection of abundant molecular species in IRC+10216 is not so much because of its unique chemical nature, but its proximity.  Furthermore, the degree of chemical activity has increased, not decreased, since the end of AGB.

Nevertheless, there are some differences in the molecular abundance of IRC+10216 and AFGL\,2688.  As shown in Figure~\ref{withirc}, the abundance of CN-bearing molecules, such as HC$_3$N and CH$_3$CN, is enhanced 
in AFGL\,2688. CN-bearing molecules have a chemical link with NH$_3$, a molecule formed near the central star. 
\citet{ngu84} found that the NH$_3$ lines in AFGL\,2688 are stronger than those in IRC+10216 by a factor of $>2$.  Therefore, it can be inferred that the injection of NH$_3$ during the AGB-to-PPN evolution enhances the production of CN-bearing molecules in AFGL\,2688 through subsequent reactions. Figure~\ref{withirc} also
indicates that Si-bearing molecules (SiS, SiO, and SiC$_2$) appear to be depleted in AFGL\,2688 with respect to IRC+10216, suggesting that the refractory species have partly condensed into dust grains.
SiS appears to be more depleted than SiO and SiC$_2$. This can be attributed to the fact that SiS has been converted into other silicon species in the outer envelope through gas-phase photochemical reactions. Detailed discussions about the chemistry of SiS and SiO in circumstellar envelopes have been given by other authors 
\citep[e.g.][]{bac97b,sch06,sch07}.  

We also found that the H$_2$CO  line in AFGL\,2688 is much stronger than
that in IRC+10216.  H$_2$CO is known to be formed through the reaction between CO and H on grain surface, and can be further transformed into methanol.
The H$_2$CO line has a broader profile (see Figure~\ref{profile}) than the other molecular lines, probably implying it originates from an extended dust region.  The detection of H$_2$CO, therefore, suggests that
grain surface chemistry may be actively processing  molecular gas in this PPN.  Based on their 3\,mm observations, \citet{luc86} found that the C$_4$H radical seems enhanced in AFGL\,2688 with respect to IRC+10216. However, this is not supported by our 1.3\,mm observations, which do not reveal C$_4$H lines 
(Figure~\ref{withirc}).  The strengths of C$_4$H lines suggest that this 
species is located at a very cool region so that the low-N transitions are 
strengthened whereas the high-N transitions are strongly suppressed.

The central-star temperature of NGC\,7027 is about 30 times higher than that of AFGL\,2688.  Figure~\ref{withngc} shows that most of neutral carbon chain molecules are destroyed during the evolution from PPN to PN, while HCO$^+$ is enhanced by a factor of $>10$. This is consistent with previous findings \citep[e.g.][]{cer11} and can be attributed to the chemistry activated by UV photons from the central star. The $^{12}$CO/$^{13}$CO line ratios in NGC\,7027 are larger than those in AFGL\,2688, which can be attributed to decreasing optical depth as the result of nebular expansion.

\section{Conclusions}

Using the {\it ARO} 12\,m and {\it SMT} 10\,m telescopes, we have carried out a molecular-line survey of the PPN  AFGL\,2688 at the 3\,mm and 1.3\,mm windows.  A total of 143 spectral lines corresponding to 32 molecular
species and isotopologues were detected. Only two faint lines remain unidentified.  Five molecules, namely C$_3$H, CH$_3$CN, H$_2$CO, H$_2$CS, and HCO$^+$, are new detections  in this object. The sensitive spectra show no evidence for the presence of C$_2$S in AFGL\,2688, contrary to the result of  \citet{park08}. We calculated the rotation temperature, the column densities and fractional abundances (or their lower-limits) of the identified molecules.

Using the observed line ratios of several molecules, we are able to determine the isotopic abundance ratios of several elements. The $^{12}$C/$^{13}$C  and $^{16}$O/$^{17}$O ratios are typical of circumstellar envelopes of evolved stars, and are considerably lower than the solar values.   The $^{14}$N/$^{15}$N ratio in AFGL\,2688
significantly differs from that in IRC+10216, probably reflecting different metallicities or/and masses of their progenitor stars.

By comparing the chemical compositions of AFGL\,2688 with those of the AGB star IRC+10216 and the PN  NGC\,7027, we have come to the conclusion that  AFGL\,2688 and NGC\,7027 have  thicker molecular envelopes
than IRC+10216.  While the chemical patterns of AFGL\,2688 and IRC+10216 are generally similar, we note that the chemistry in NGC 7027 is different, probably as the result of a drastically different radiation environment and photochemistry.   Nevertheless, we do note that there are slight differences between the molecular line strengths in and AFGL\,2688 and IRC+10216,  providing evidence for chemical evolution
during the AGB-to-PPN transition. Specially, CN-containing molecules are enhanced and refractory molecules are depleted as the result of stellar evolution.  

While we are able to come to some useful conclusions on chemical evolution in the circumstellar environment, we do wish to note that the origin of chemical diversity  in circumstellar
envelopes can be complicted by a number of factors.   The difference in molecular abundance can be the result of stellar evolution, metallicities, initial masses, mass-loss processes, dust properties, shock processes, morphological structures, interactions with the surrounding interstellar medium, just to name a few. In order to understand the roles of these various factors in circumstellar chemistry, it would be necessary to  perform high-sensitivity spectral line survey of a large sample of evolved stars over wide frequency ranges.   High angular resolution mapping of a large variety of lines would also be useful.  We hope that this paper will provide the impetus of future work using larger facilities such as ALMA.

\acknowledgments

 We thank the anonymous referee for helpful comments.
The 12\,m telescope and the {\it SMT} are operated by the Arizona Radio Observatory ({\it ARO}), Steward Observatory, University of Arizona.  We are grateful to the {\it ARO} staff for their help during the observing run.
Financial support for this work was provided by the Research Grants Council of the Hong Kong under grants HKU7073/11P and HKU7027/11P.  Dinh Van Trung acknowledged the financial support from 
Vietnam National Foundation for Science and Technology 
(NAFOSTED) under contract 103.08-2010.26.

\begin{figure*}
\plotone{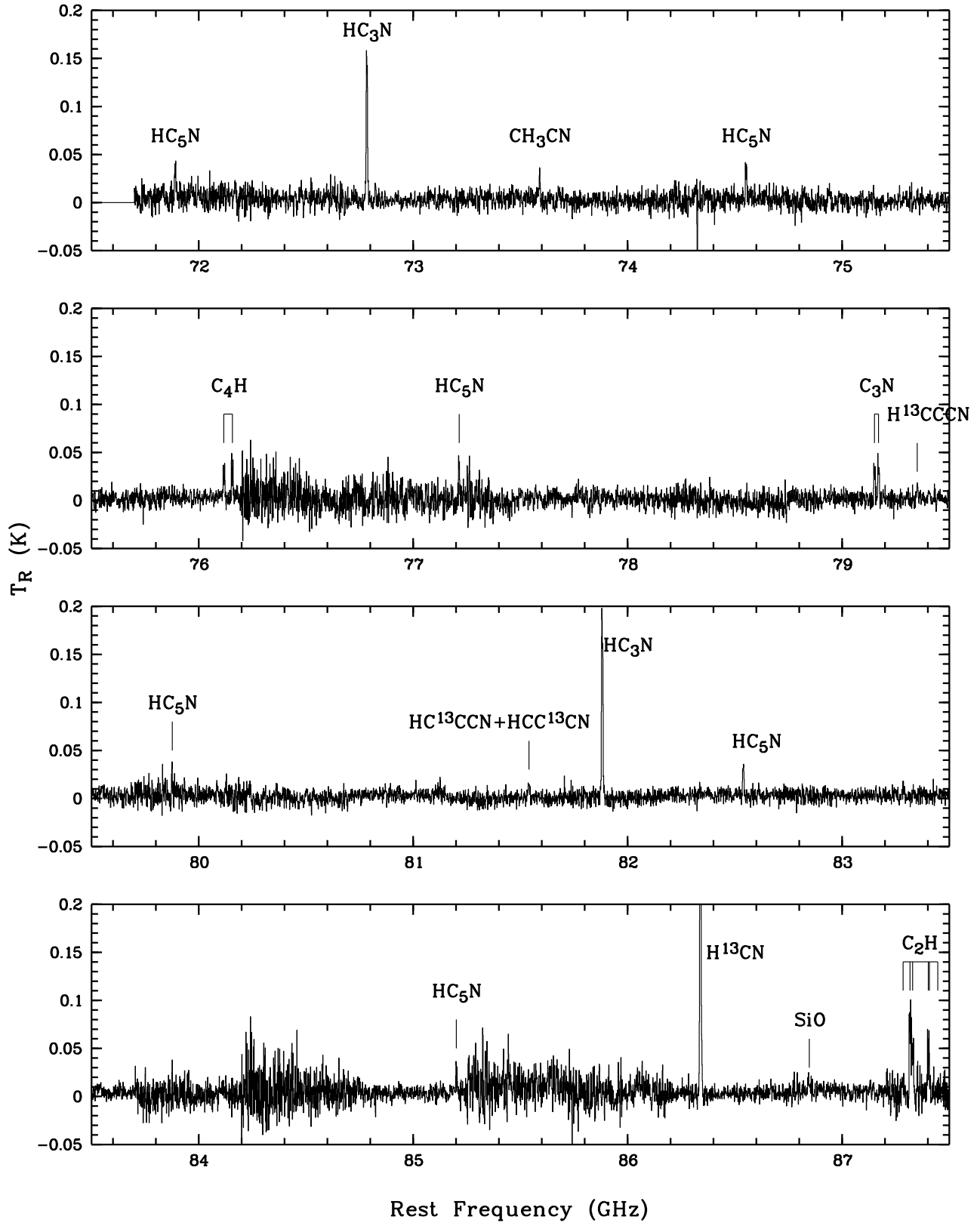}
\caption{Spectrum of AFGL\,2688 in the frequency range
71--111 and 157--160\,GHz obtained with the {\it ARO} 12\,m telescope. 
The spectra have been smoothed to a resolution of 1\,MHz.
(The data used to create this figure are available in the online journal.)
}
\label{spe_12m}
\end{figure*}

\addtocounter{figure}{-1}
\begin{figure*}
\plotone{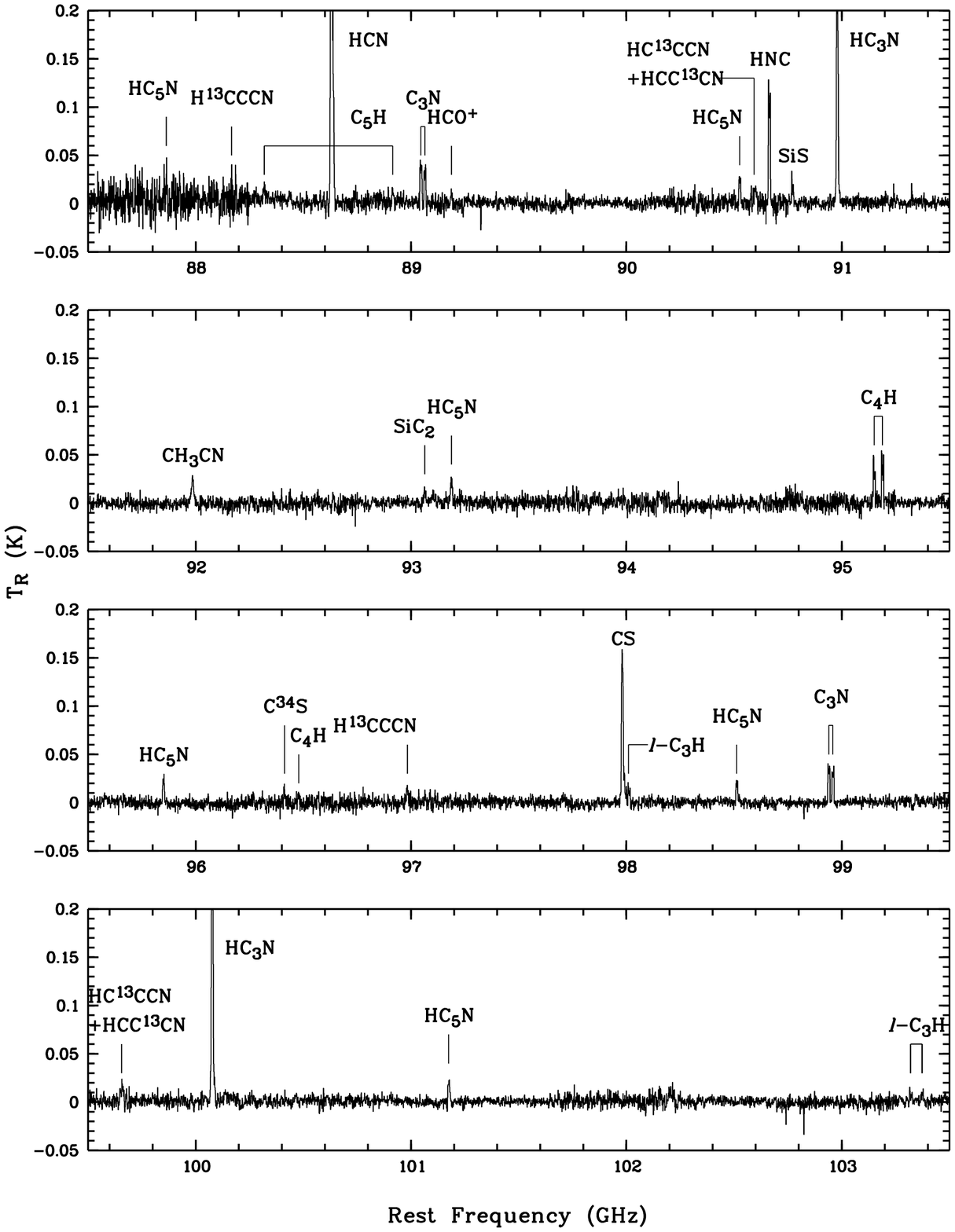}
\caption{continued.}
\end{figure*}

\addtocounter{figure}{-1}
\begin{figure*}
\plotone{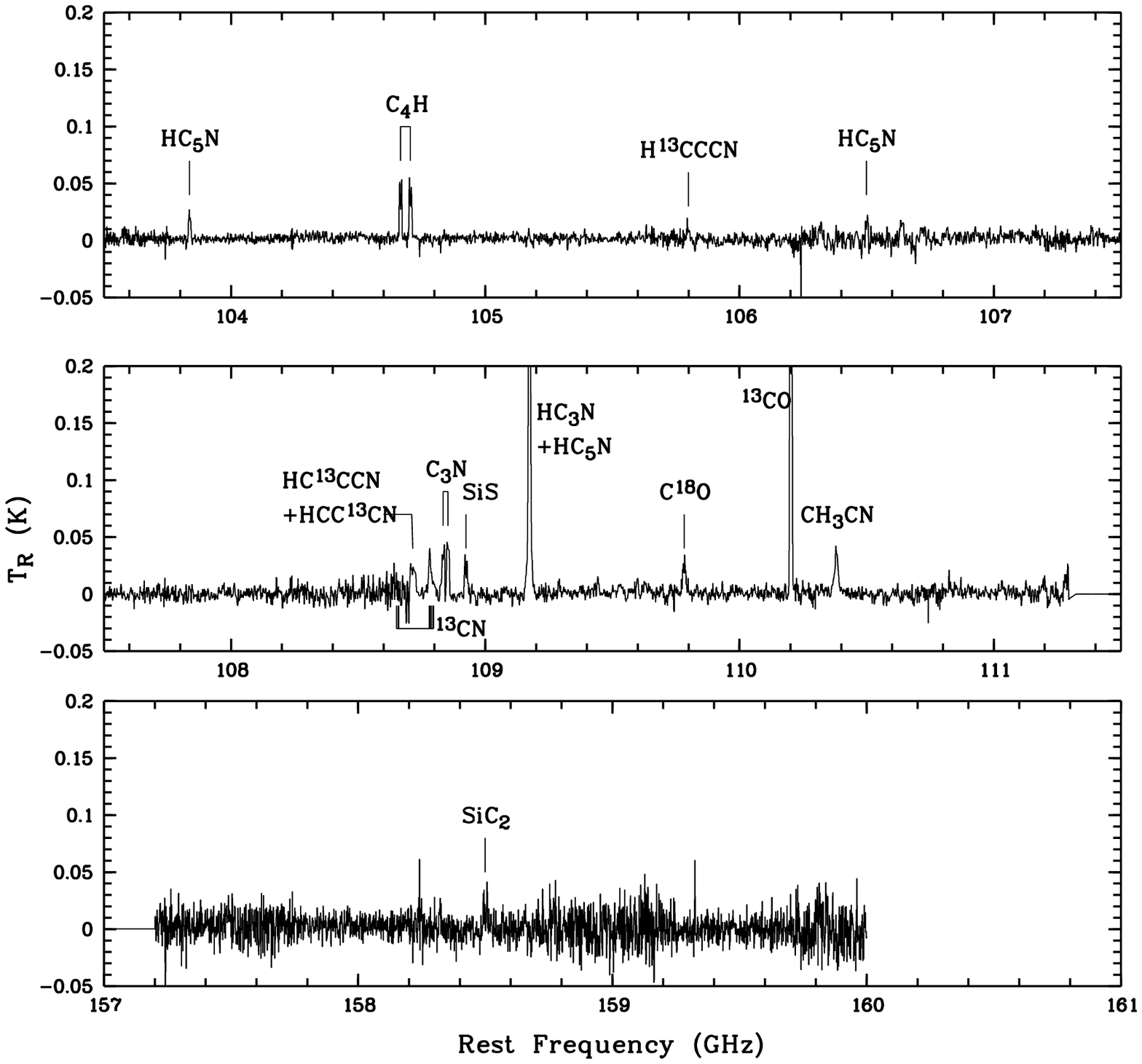}
\caption{continued.}
\end{figure*}

\begin{figure*}
\plotone{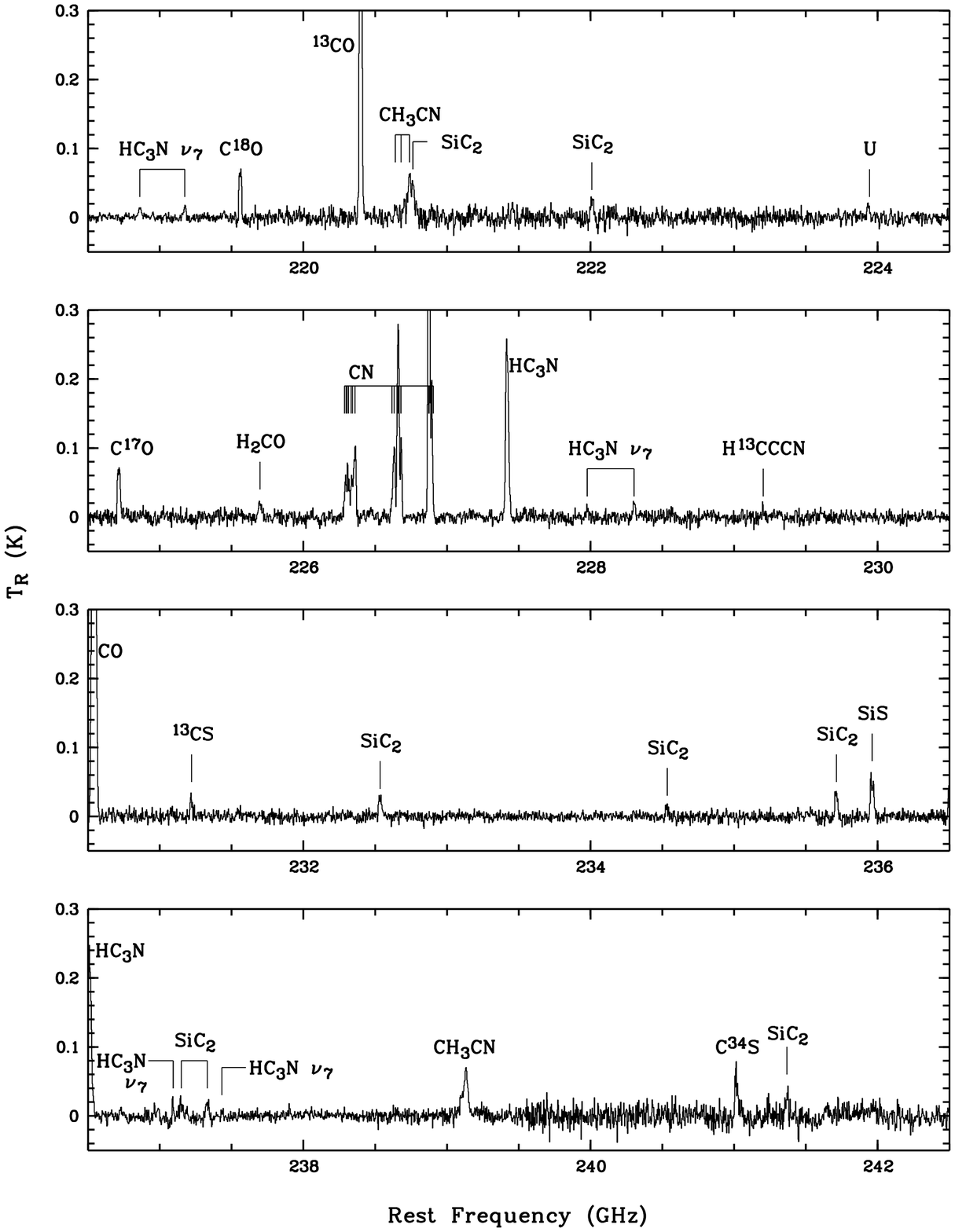}
\caption{Spectrum of AFGL\,2688 in the frequency ranges
218--267\,GHz obtained with the {\it SMT}.
The spectra have been smoothed to a resolution of 3\,MHz.
(The data used to create this figure are available in the online journal.)
}
\label{spe_smt}
\end{figure*}

\addtocounter{figure}{-1}
\begin{figure*}
\plotone{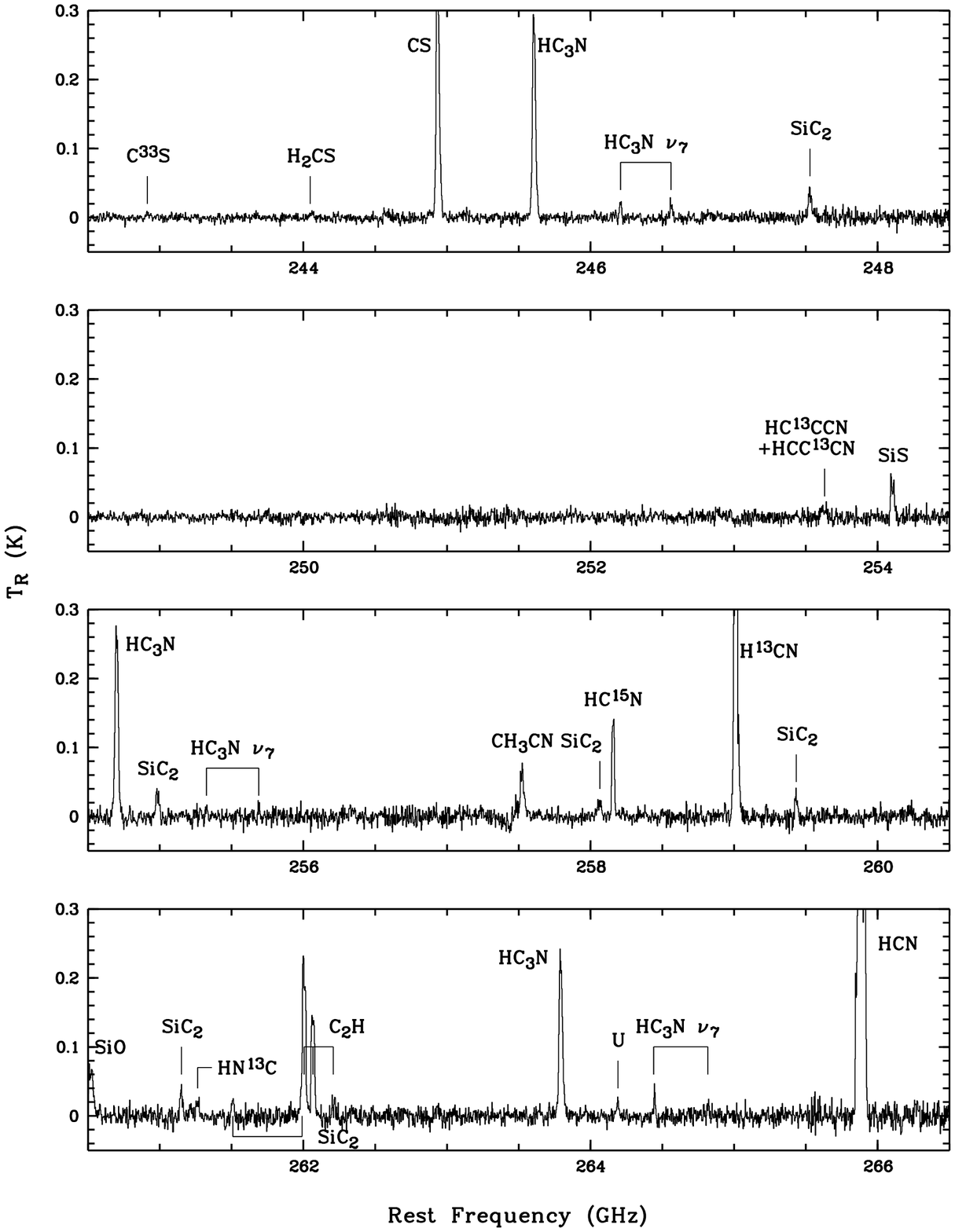}
\caption{continued.}
\end{figure*}

\begin{figure*}
\center
\epsfig{file=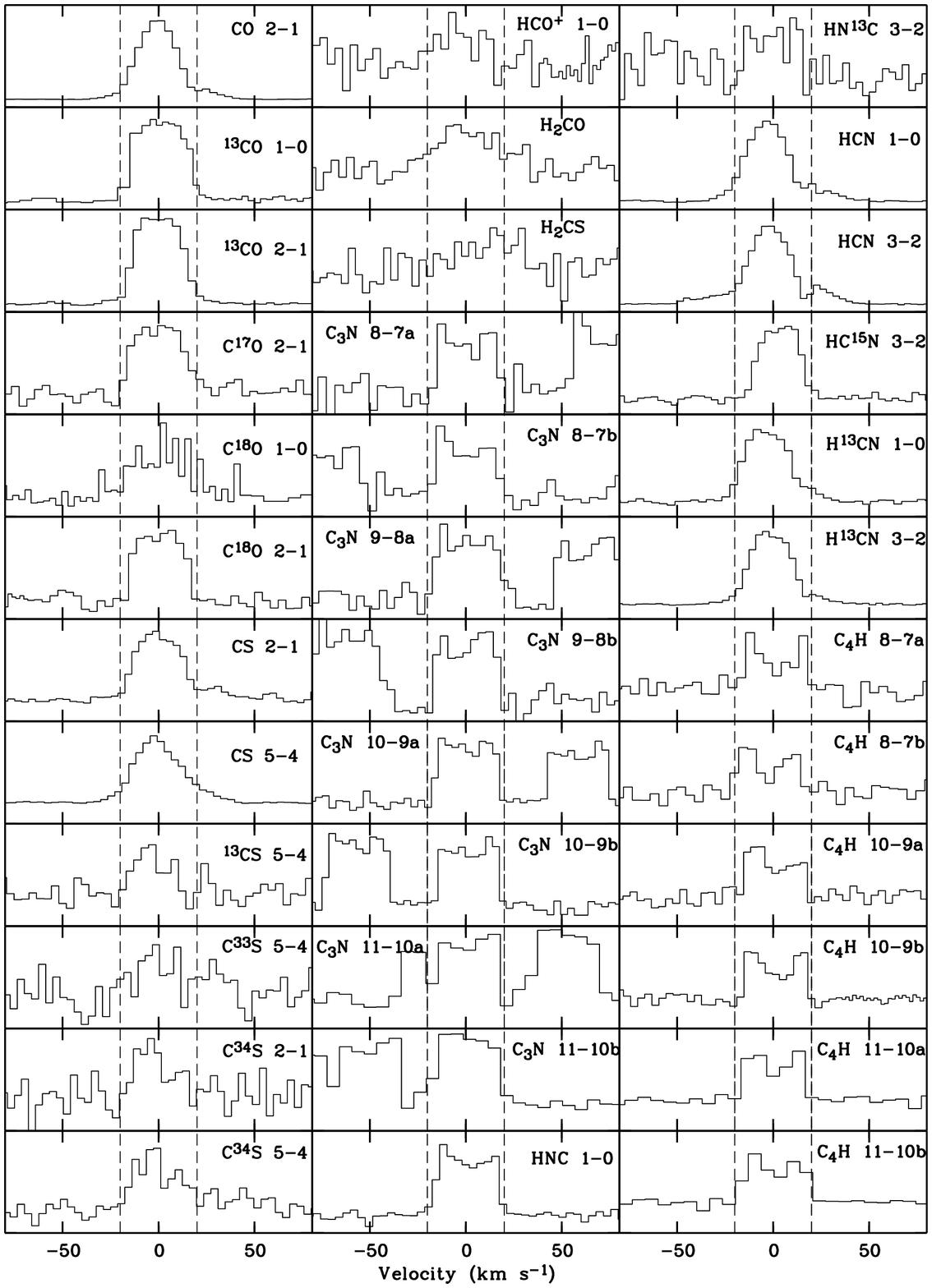, height=22cm}
\caption{Line profiles. The vertical dashed lines mark the positions of 
$v=\pm20$\,km\,s$^{-1}$.  The wavelengths of CH$_3$CN hyperfine components are marked by arrows.
The intensity scale is arbitrary.}
\label{profile}
\end{figure*}

\addtocounter{figure}{-1}
\begin{figure*}
\centering
\epsfig{file=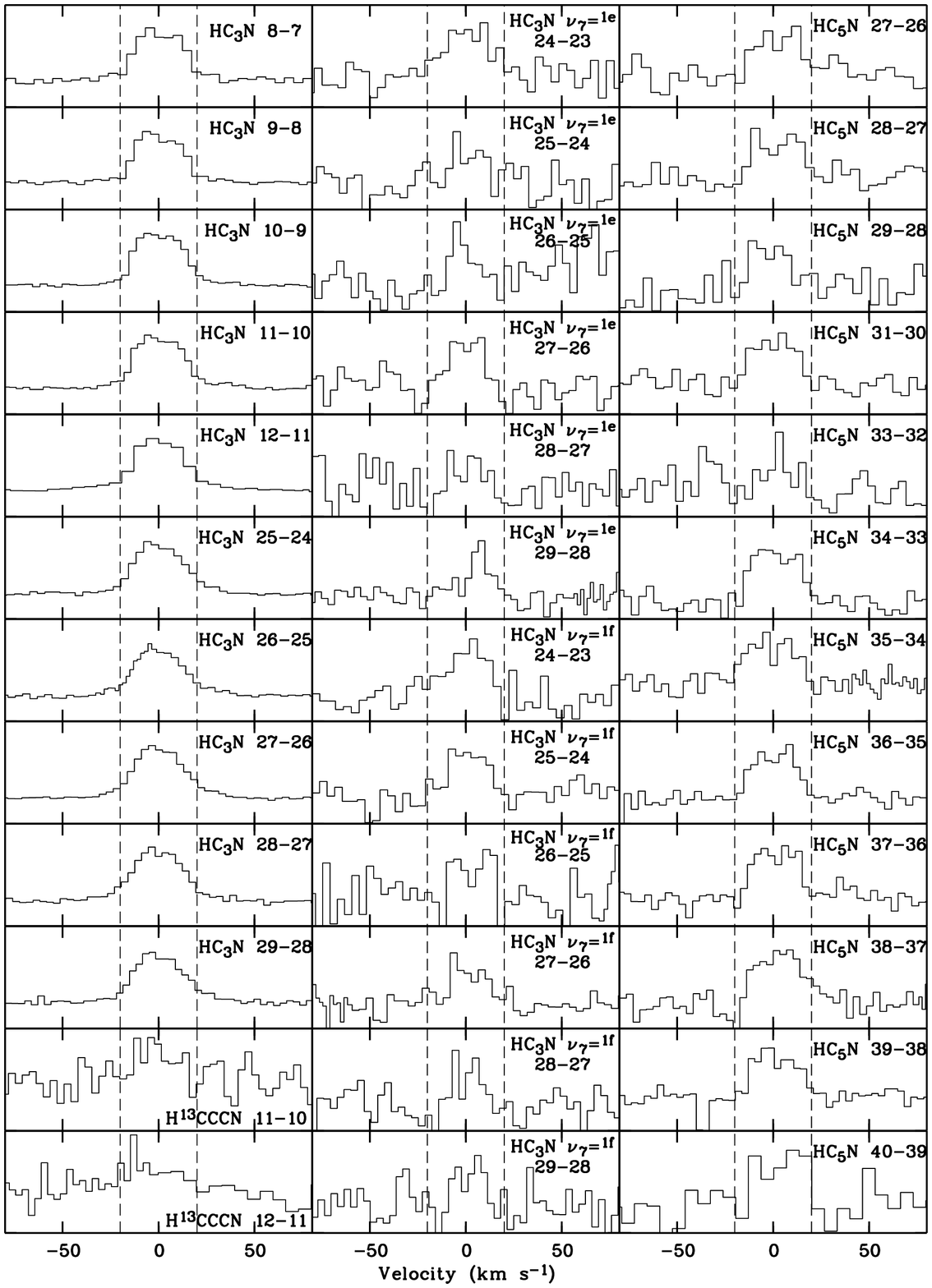,
height=22cm, }
\caption{continued.}
\end{figure*}

\addtocounter{figure}{-1}
\begin{figure*}
\epsfig{file=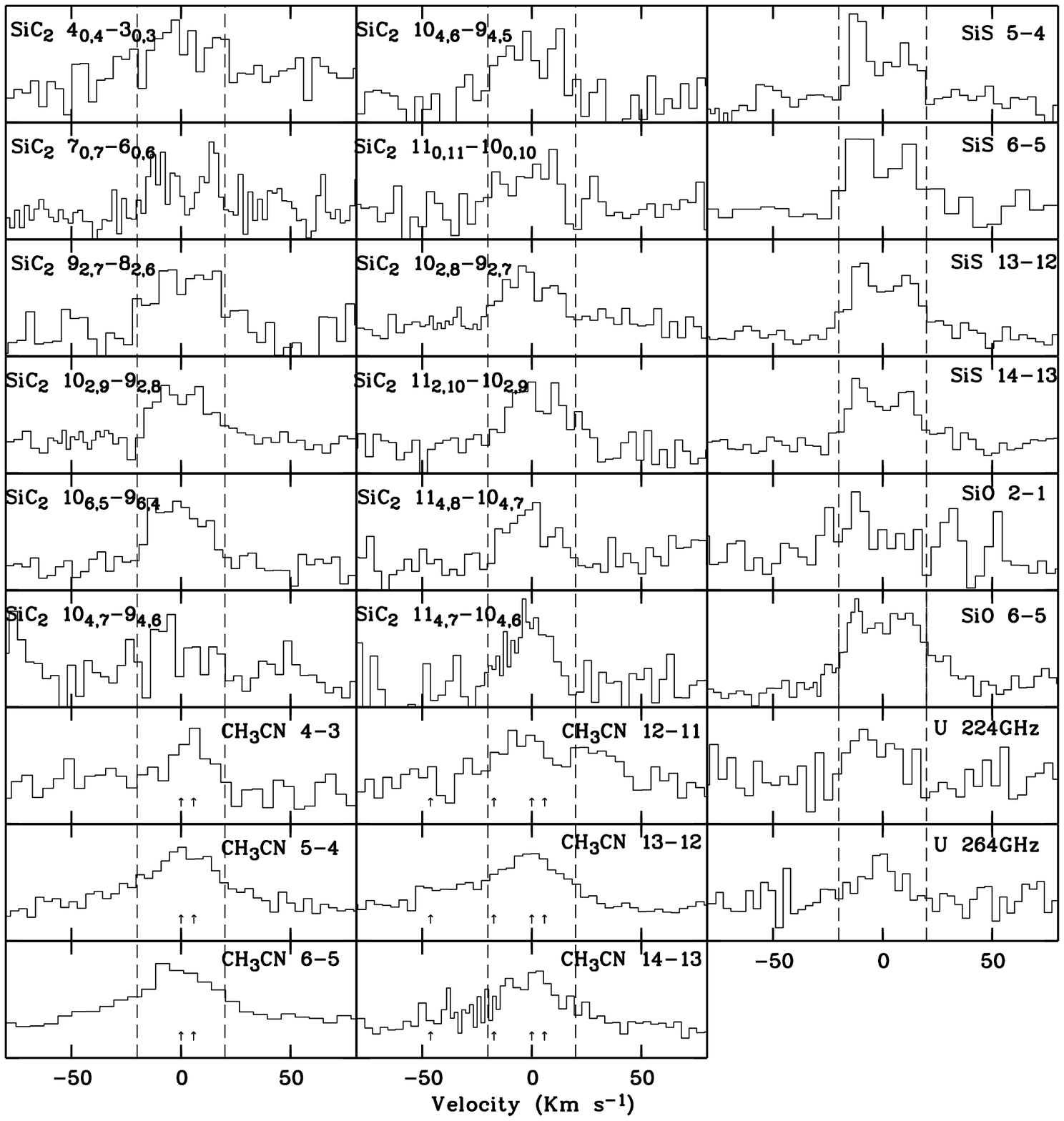,
height=17cm, }
\caption{continued.}
\end{figure*}

\begin{figure*}
\epsfig{file=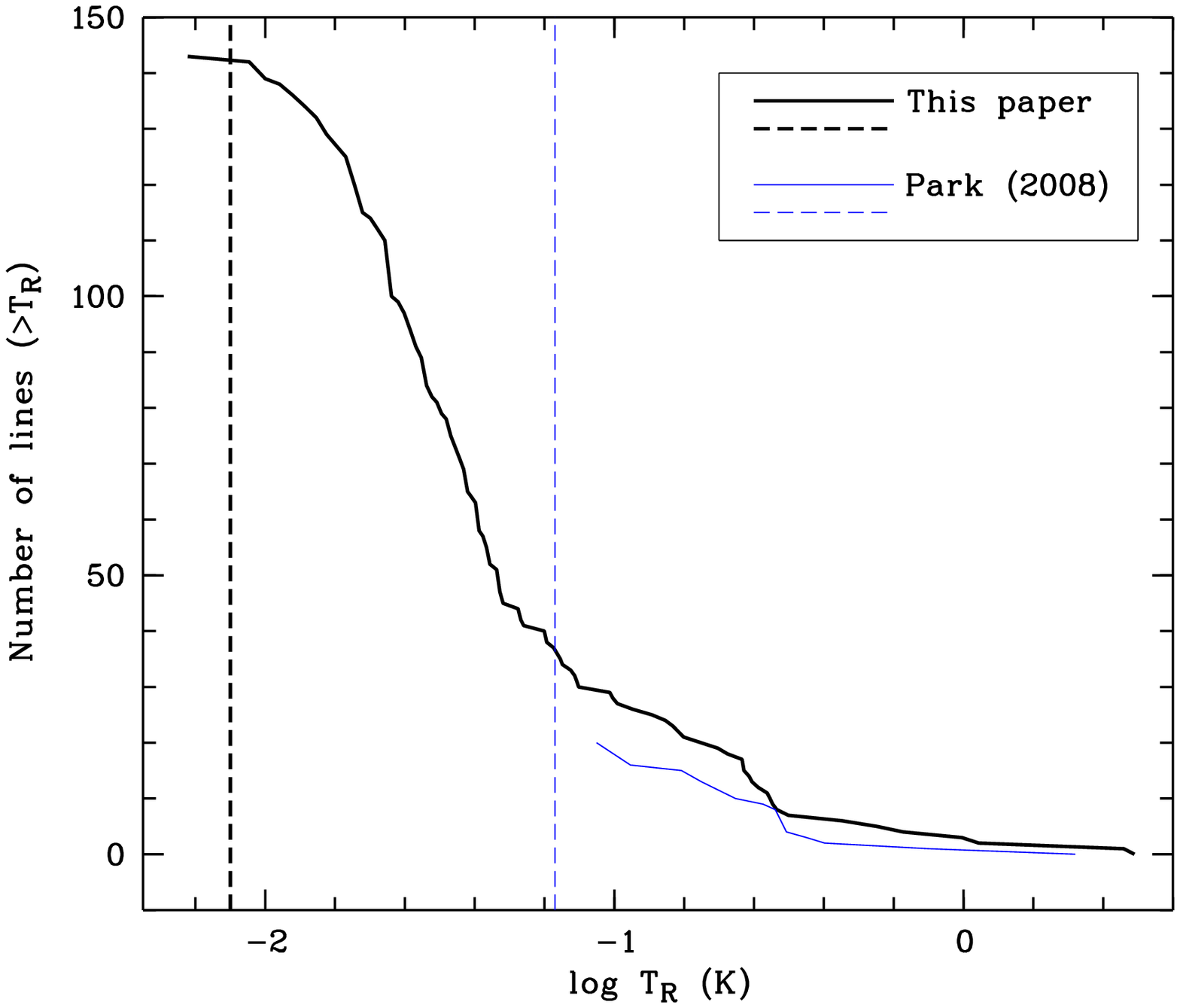,
height=13cm}
\caption{Cumulative number of observed lines in this survey exceeding a given
$T_{\rm R}$ value. The results from \citet{park08} are overplotted for 
comparison.  The dashed lines mark the typical sensitivities.}
\label{cum}
\end{figure*}

\begin{figure*}
\epsfig{file=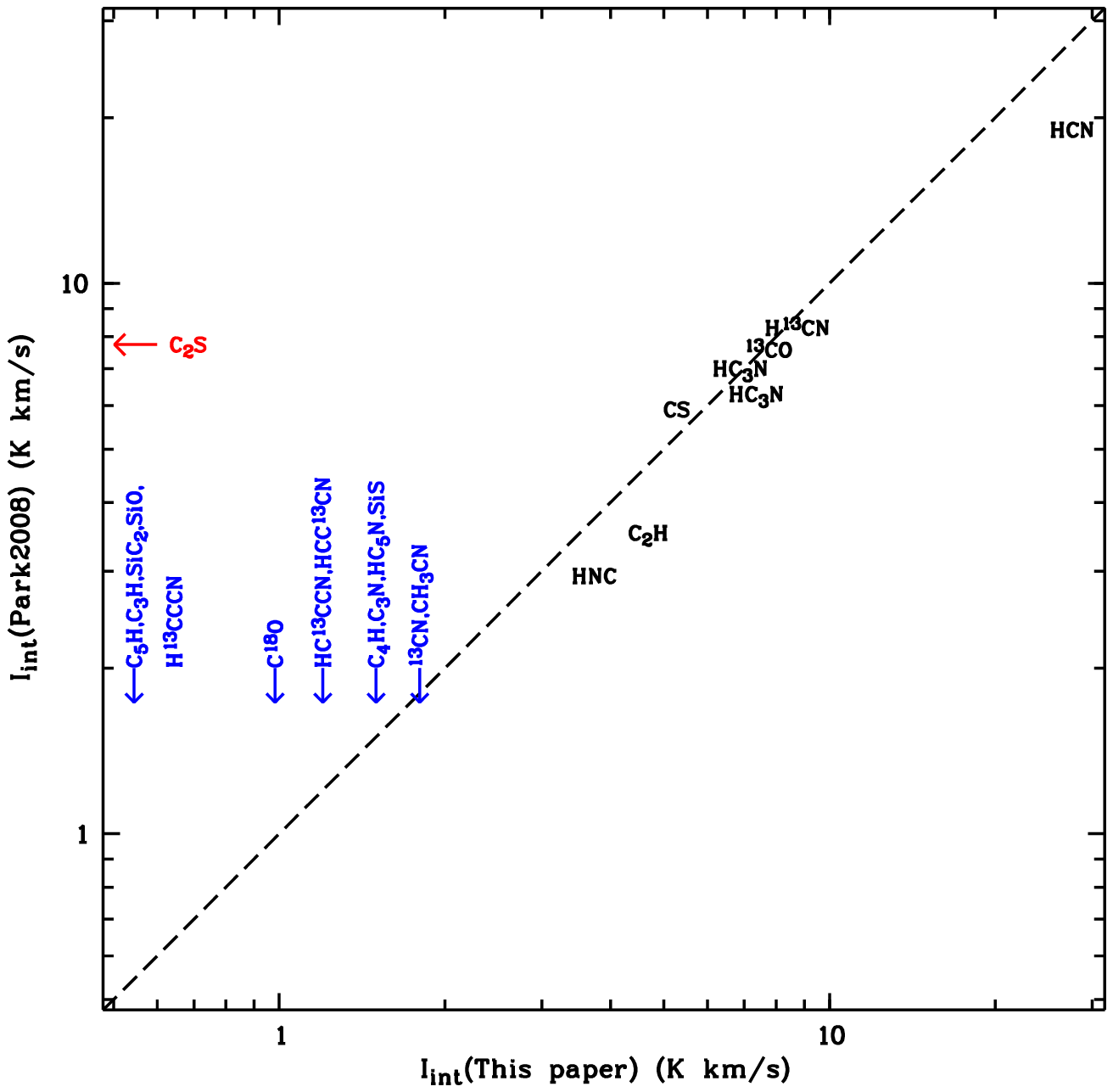,
height=15cm}
\caption{Comparison the strengths ($\int T_{\rm R} dv$) of molecular lines 
at 85--111\,GHz detected in this paper and those by
\citet{park08}. The diagonal line is a $y=x$ plot. The species
indicated by blue color are those not detected by \citet{park08}.
}
\label{park}

\end{figure*}

\begin{figure*}
\epsfig{file=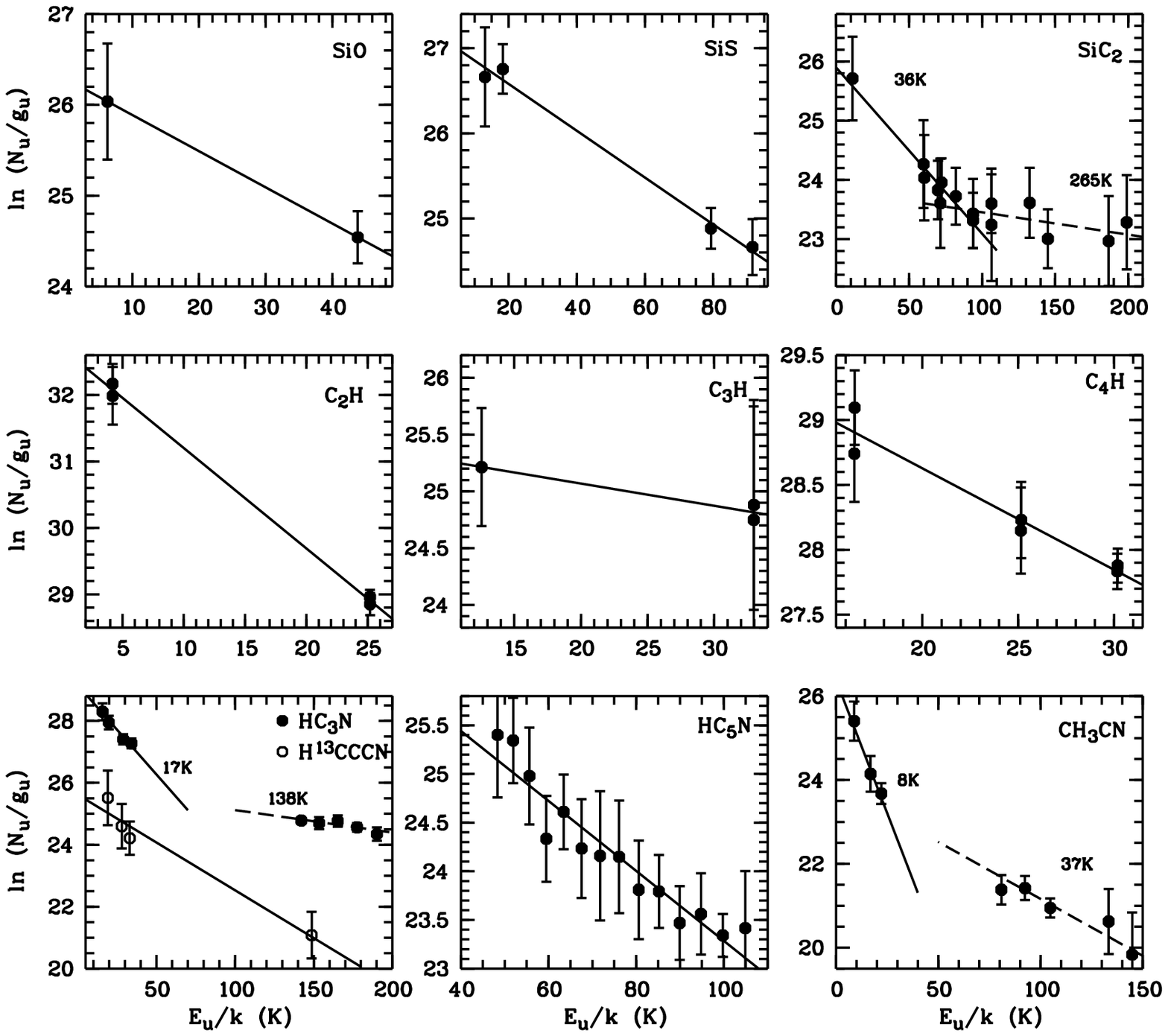,
height=15cm}
\caption{Rotational diagrams for the detected species in AFGL\,2688.
}
\label{dia}
\end{figure*}

\begin{figure*}
\epsfig{file=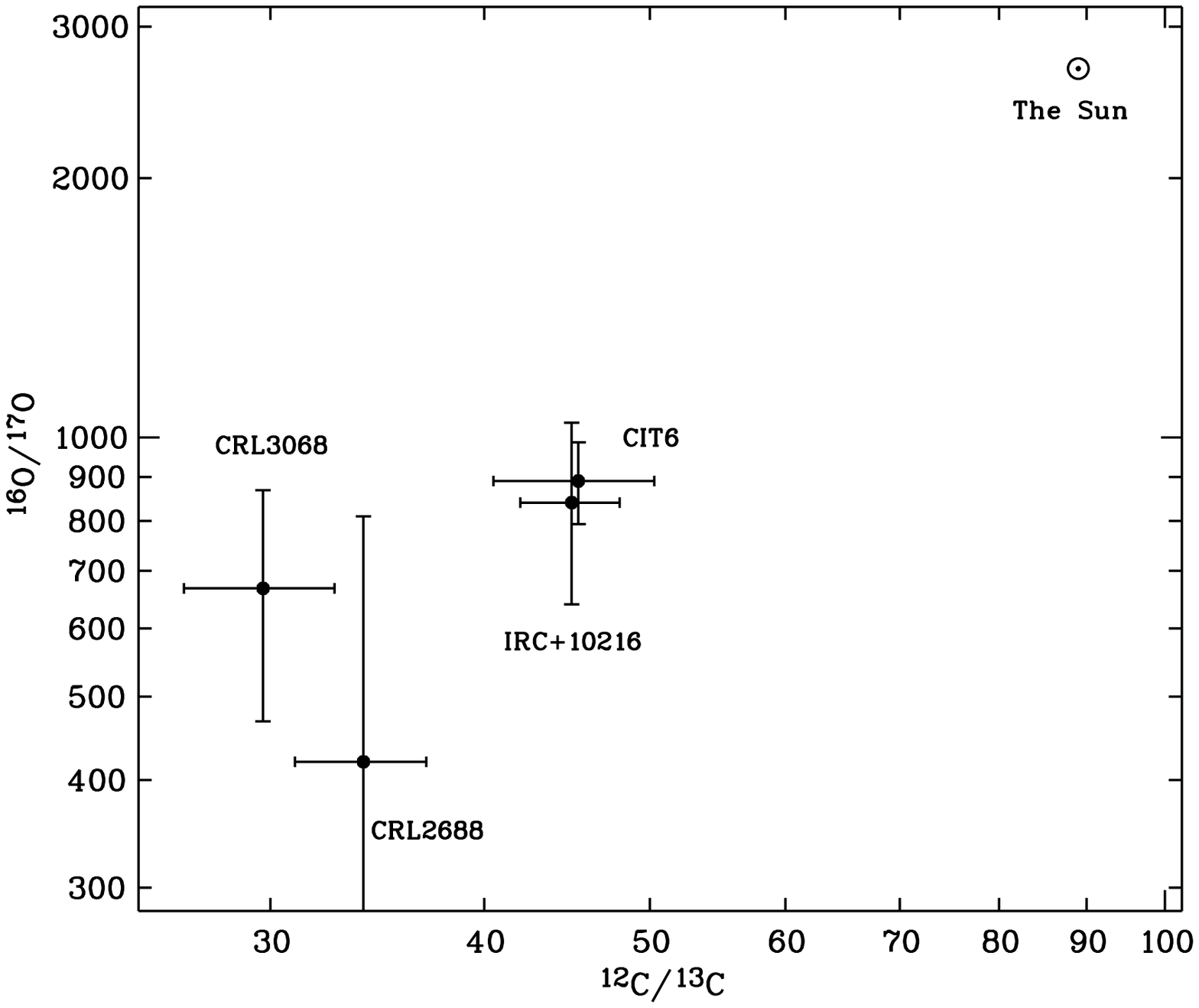,
height=13cm}
\caption{The $^{16}$O/$^{17}$O versus $^{12}$C/$^{13}$C abundance ratios
for the objects in our long-term line-survey project.
}
\label{isoto}
\end{figure*}

\begin{figure*}
\epsfig{file=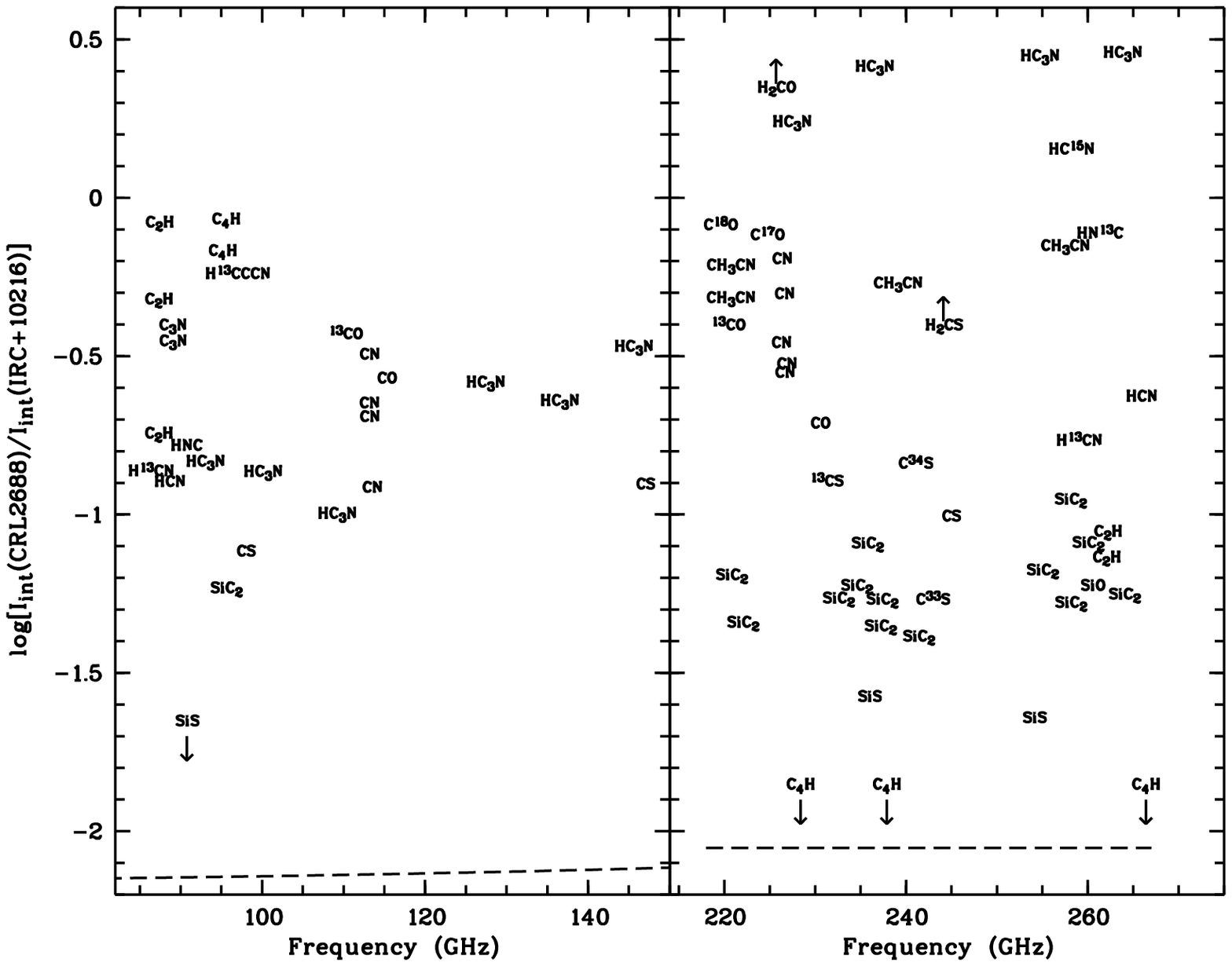,
height=13cm}
\caption{Integrated strength ratio of the lines 
3\,mm (left panel) and 1.3\,mm (right panel) detected in AFGL\,2688
and those detected in IRC+10216. Note that the 3\,mm data are taken
from \citet{park08} except those of SiC$_2$, C$_4$H, C$_3$N, and 
H$^{13}$CCCN taken from \citet{luc86}.
The dashed lines represent the predicted values under the assumption
that the two objects have equal line fluxes at an equal distance.
The assumed values for $\theta_s$ are 30\arcsec\ and 20\arcsec\ for IRC+10216 and AFGL 2688 respectively.
}
\label{withirc}
\end{figure*}

\begin{figure*}
\epsfig{file=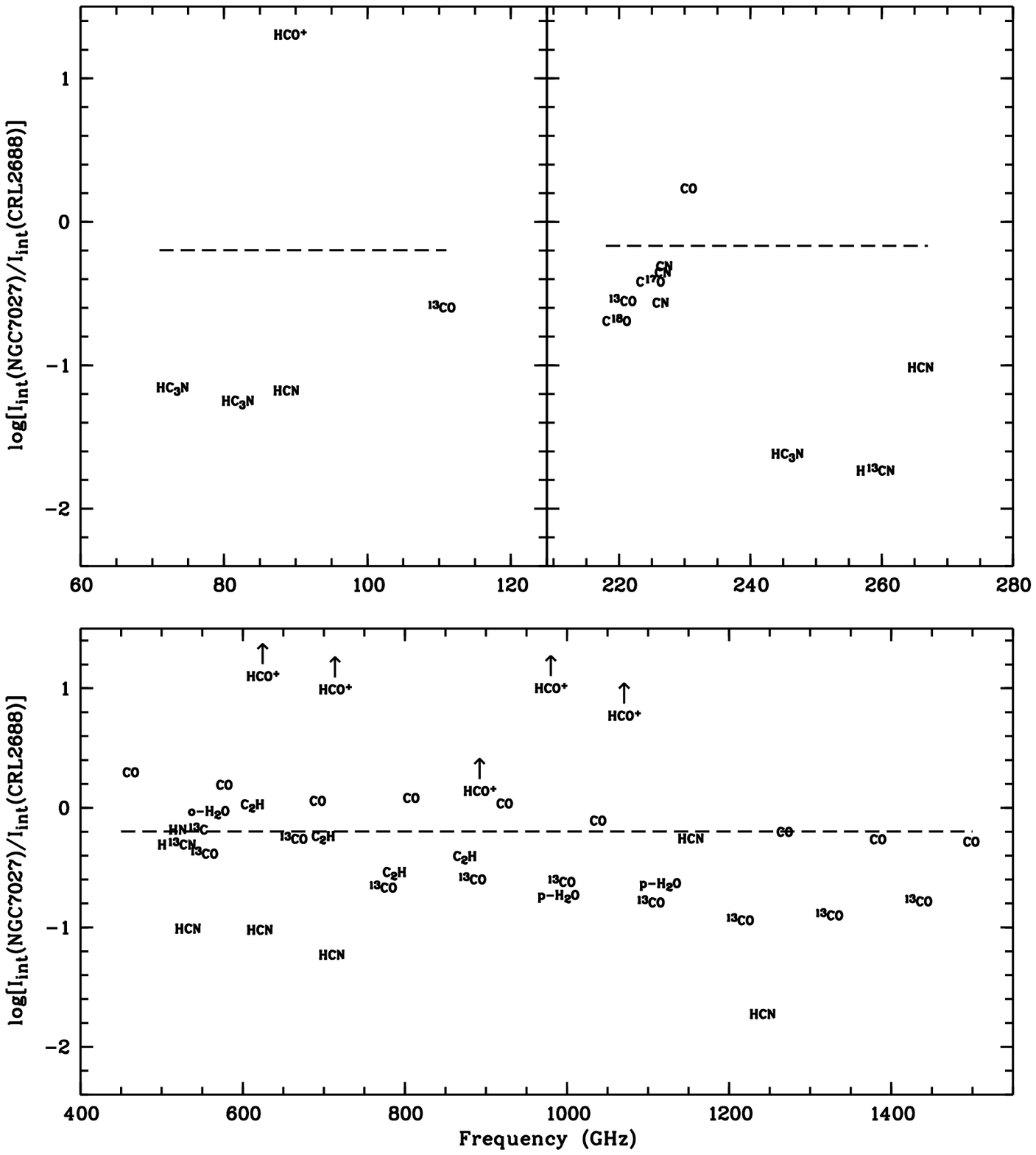,
height=18cm}
\caption{Integrated strength ratio of the 3\,mm (left panel), 1.3\,mm (right
panel), and sub-mm (down panel) lines detected in NGC\,7027 and those detected in AFGL\,2688. 
Note that the sub-mm data are taken from \citet{wes10}.
The dashed lines represent the predicted values under the assumption
that the two objects have equal line fluxes at an equal distance.  The assumed values for $\theta_s$ are 13.2\arcsec\ and 20\arcsec\ for NGC 7027 and AFGL 2688 respectively.
}
\label{withngc}
\end{figure*}

\begin{deluxetable}{lllllll}
\tablecaption{Molecular transitions detected in  AFGL~2688. The
frequencies given here are rest-frequencies from catalogs.
\label{crl}}
\tabletypesize{\scriptsize}
\tablewidth{0pt}
\tablehead{
\colhead{Frequency}& \colhead{Species} & \colhead{Transition} &  \colhead{rms} & \colhead{$T_{\rm R}$} &\colhead{$\int T_{\rm R}$d$v$} &\colhead{Note}\\
\colhead{(MHz)} & &\colhead{(upper--lower)} &  \colhead{(mK)} &\colhead{(K)} & \colhead{(K~km/s)} & \\
}
\startdata
  226287.4&  CN          & N(J,F)=2(3/2,1/2)--1(3/2,1/2)     &6.2  &   0.078&    2.860 &   $a$\\
  226298.9&  CN          & N(J,F)=2(3/2,1/2)--1(3/2,3/2)     &---  &   ---  &    ---   &   $a$         \\
  226303.0&  CN          & N(J,F)=2(3/2,3/2)--1(3/2,1/2)     &---  &   ---  &    ---   &   $a$  \\
  226314.5&  CN          & N(J,F)=2(3/2,3/2)--1(3/2,3/2)     &---  &   ---  &    ---   &   $a$ \\
  226332.5&  CN          & N(J,F)=2(3/2,3/2)--1(3/2,5/2)     &6.2  &   0.102&    3.693 &   $a$\\
  226341.9&  CN          & N(J,F)=2(3/2,5/2)--1(3/2,3/2)     &---  &   ---  &    ---   &  $a$      \\
  226359.9&  CN          & N(J,F)=2(3/2,5/2)--1(3/2,5/2)     &---  &   ---  &    ---   &   $a$\\
  226616.6&  CN          & N(J,F)=2(3/2,1/2)--1(1/2,3/2)     &5.4  &   0.099&    2.756 &   $a$\\
  226632.2&  CN          & N(J,F)=2(3/2,3/2)--1(1/2,3/2)     &---  &   ---  &    ---   &   $a$\\
  226659.6&  CN          & N(J,F)=2(3/2,5/2)--1(1/2,3/2)     &5.4  &   0.279&    6.460 &   $a$\\
  226663.7&  CN          & N(J,F)=2(3/2,1/2)--1(1/2,1/2)     &---  &   ---  &    ---   &   $a$ \\
  226679.3&  CN          & N(J,F)=2(3/2,3/2)--1(1/2,1/2)     &4.5  &   0.113&    1.917 &   $a$\\
  226874.2&  CN          & N(J,F)=2(5/2,5/2)--1(3/2,3/2)     &4.5  &   0.566&   12.804 &   $a$\\
  226874.8&  CN          & N(J,F)=2(5/2,7/2)--1(3/2,5/2)     &---  &   ---  &    ---   &   $a$\\
  226875.9&  CN          & N(J,F)=2(5/2,3/2)--1(3/2,1/2)     &---  &   ---  &    ---   &   $a$\\
  226887.4&  CN          & N(J,F)=2(5/2,3/2)--1(3/2,3/2)     &---  &   ---  &    ---   &   $a$\\
  226892.1&  CN          & N(J,F)=2(5/2,5/2)--1(3/2,5/2)     &4.5  &   0.198&    3.007 &   $a$ \\
  226905.4&  CN          & N(J,F)=2(5/2,3/2)--1(3/2,5/2)     &---  &   ---  &    ---   &   $a$\\
  108651.3&  $^{13}$CN   & 1/2--1/2 F=2--1 F$_1$=0 F$_2$=1--0&6.0  &   0.026&    0.352:&    $a$\\
  108657.6&  $^{13}$CN   & 1/2--1/2 F=2--2 F$_1$=1 F$_2$=1--1&---  &   ---  &    ---   &    $a$\\
  108658.9&  $^{13}$CN   & 1/2--1/2 F=1--2 F$_1$=1 F$_2$=1--1& --- &   ---  &    ---   &    $a$\\
  108780.2&  $^{13}$CN   & 3/2--1/2 F=3--2 F$_1$=1 F$_2$=2--1&6.0  &   0.038&    1.702 &    $a$\\
  108782.4&  $^{13}$CN   & 3/2--1/2 F=2--1 F$_1$=1 F$_2$=2--1& --- &   ---  &    ---   &    $a$\\
  108787.0&  $^{13}$CN   & 3/2--1/2 F=1--0 F$_1$=1 F$_2$=2--1& --- &   ---  &    ---   &    $a$\\
  108793.8&  $^{13}$CN   & 3/2--1/2 F=1--1                   & --- &   ---  &    ---   &    $a$   \\
  108796.4&  $^{13}$CN   & 3/2--1/2 F=2--2                   & --- &   ---  &    ---   &    $a$   \\
  230538.0&  CO          & J=2--1                            &5.4  &   3.087&   83.055 &\\
  110201.4&  $^{13}$CO   & J=1--0                            &3.9  &   0.235&    7.702 &\\
  220398.7&  $^{13}$CO   & J=2--1                            &7.3  &   0.671&   20.398 &\\
  224714.4&  C$^{17}$O   & J=2--1                            &5.8  &   0.071&    2.248 &\\
  109782.1&  C$^{18}$O   & J=1--0                            &3.8  &   0.032&    0.983 &\\
  219560.4&  C$^{18}$O   & J=2--1                            &4.9  &   0.070&    2.015 &\\
   97981.0&  CS          &J=2--1                             &3.5  &   0.158&    5.252 &\\
  244935.6&  CS          &J=5--4                             &4.5  &   0.449&   13.162 &        \\
  231221.0&  $^{13}$CS   &J=5--4                             &5.2  &   0.034&    0.720 &\\
  242913.6&  C$^{33}$S   &J=5--4                             &3.5  &   0.009&    0.162:&\\
   96412.9&  C$^{34}$S   &J=2--1                             &3.5  &   0.015&    0.385 &\\
  241016.1&  C$^{34}$S   &J=5--4                             &10.0 &   0.079&    1.950 &\\
   86847.0&  SiO         &J=2--1                             &4.9  &   0.023&    0.397 &\\
  260518.0&  SiO         &J=6--5                             &7.2  &   0.075&    3.013 &       \\
   90771.5&  SiS         &J=5--4                             &6.6  &   0.034&    0.682 &\\
  108924.3&  SiS         &J=6--5                             &3.2  &   0.033&    1.310 &\\
  235961.1&  SiS         &J=13--12                           &5.1  &   0.064&    1.931 &\\
  254102.9&  SiS         &J=14--13                           &6.6  &   0.063&    1.957 &\\
   87284.6&  C$_2$H      & N=1--0 J,F=3/2,1--1/2,1           &9.7  &   0.020&    0.580:&      \\
   87317.1&  C$_2$H      & N=1--0 J,F=3/2,2--1/2,1           &9.7  &   0.097&    4.670 &    $a$\\
   87328.7&  C$_2$H      & N=1--0 J,F=3/2,1--1/2,0           &---  &   ---  &    ---   &    $a$\\
   87402.1&  C$_2$H      & N=1--0 J,F=1/2,1--1/2,1           &9.7  &   0.067&    1.820 &    $a$       \\
   87407.2&  C$_2$H      & N=1--0 J,F=1/2,0--1/2,1           &---  &   ---  &    ---   &    $a$   \\
   87446.4&  C$_2$H      & N=1--0 J,F=1/2,1--1/2,0           &9.7  &   0.019&    0.510:&\\
  262005.3&  C$_2$H      &N=3--2 J=7/2--5/2                  &8.2  &   0.232&    7.621 &      $b$\\
  262066.1&  C$_2$H      &N=3--2 J=5/2--3/2                  &8.2  &   0.147&    4.630 &\\
  262208.4&  C$_2$H      &N=3--2 J=5/2--5/2                  &8.2  &   0.030&    0.540:&\\
   88631.6&  HCN         & J=1--0                            &4.9  &   0.990&   29.090 &\\
  265886.4&  HCN         & J=3--2                            &10.6 &   2.878&   84.870 &    $a$\\
  265852.8&  HCN         & $\nu_2$=1$^{1e}$ J=3--2           &---  &   ---  &    ---   &    $a$       \\
  258154.7&  HC$^{15}$N  & J=3--2                            &5.5  &   0.140&    3.593 &       \\
  86340.2&  H$^{13}$CN  & J=1--0                             &5.0  &   0.315&    8.710 &\\
  259011.8&  H$^{13}$CN  & J=3--2                            &6.9  &   1.106&   30.313 &\\
   90663.6&  HNC         & J=1--0                            &4.9  &   0.128&    3.720 &\\
  261263.4&  HN$^{13}$C  & J=3--2                            &7.5  &   0.027&    0.618 &\\
   93063.6&  SiC$_2$     & 4$_{0,4}$--3$_{0,3}$              &4.0  &   0.017&    0.410 &       \\
  158499.2&  SiC$_2$     & 7$_{0,7}$--6$_{0,6}$              &8.4  &   0.034:&    0.320: &\\
  220773.7&  SiC$_2$     & 10$_{0,10}$--9$_{0,9}$            &7.4  &   0.054&    1.333 &\\
  222009.4&  SiC$_2$     & 9$_{2,7}$--8$_{2,6}$              &7.2  &   0.030&    0.923 &\\
  232534.1&  SiC$_2$     & 10$_{2,9}$--9$_{2,8}$             &5.1  &   0.031&    0.930 &\\
  234534.0&  SiC$_2$     & 10$_{8,2}$--9$_{8,1}$             &4.8  &   0.019&    0.298 &   $a$\\
  234534.0&  SiC$_2$     & 10$_{8,3}$--9$_{8,2}$             & --- &   ---  &    ---   &   $a$\\
  235713.0&  SiC$_2$     & 10$_{6,5}$--9$_{6,4}$             &5.9  &   0.036&    1.016 &   $a$  \\
  235713.1&  SiC$_2$     & 10$_{6,4}$--9$_{6,3}$             & --- &   ---  &    ---   &   $a$  \\
  237150.0&  SiC$_2$     & 10$_{4,7}$--9$_{4,6}$             &4.5  &   0.029&    0.505 &\\
  237331.3&  SiC$_2$     & 10$_{4,6}$--9$_{4,5}$             &4.5  &   0.023&    0.567 &\\
  241367.7&  SiC$_2$     & 11$_{0,11}$--10$_{0,10}$          &10.8 &   0.043&    0.903 &\\
  247529.1&  SiC$_2$     & 10$_{2,8}$--9$_{2,7}$             &6.0  &   0.044&    1.177 &\\
  254981.5&  SiC$_2$     & 11$_{2,10}$--10$_{2,9}$           &6.4  &   0.040&    1.091 &\\
  258065.0&  SiC$_2$     & 11$_{8,3}$--10$_{8,2}$            &6.1  &   0.023&    0.716 &   $a$\\
  258065.0&  SiC$_2$     & 11$_{8,4}$--10$_{8,3}$            &---  &   ---  &    ---   &   $a$\\
  259433.3&  SiC$_2$     & 11$_{6,6}$--10$_{6,5}$            &6.8  &   0.041&    0.800 &   $a$  \\
  259433.3&  SiC$_2$     & 11$_{6,5}$--10$_{6,4}$            &---  &   ---  &    ---   &   $a$ \\
  261150.7&  SiC$_2$     & 11$_{4,8}$--10$_{4,7}$            &7.6  &   0.046&    0.902 &\\   
  261509.3&  SiC$_2$     & 11$_{4,7}$--10$_{4,6}$            &7.9  &   0.025&    0.631 &\\
  261990.7&  SiC$_2$     & 12$_{0,12}$--11$_{0,11}$          &---  &   ---  &    ---   &   $b$ \\
   98011.6&  C$_3$H      & $^2\Pi_{1/2}$ J=9/2--7/2 F=5--4 f &3.3  &   0.019&   0.545  &  $a$ \\
   98012.5&  C$_3$H      & $^2\Pi_{1/2}$ J=9/2--7/2 F=4--3 f &---  &   ---  &   ---    &  $a$ \\
  103319.3&  C$_3$H      & $^2\Pi_{3/2}$ J=9/2--7/2 F=5--4 f &4.0  &   0.012&   0.367  &  $a$ \\
  103319.8&  C$_3$H      & $^2\Pi_{3/2}$ J=9/2--7/2 F=4--3 f &---  &   ---  &   ---    &  $a$ \\
  103372.5&  C$_3$H      & $^2\Pi_{3/2}$ J=9/2--7/2 F=5--4 e &4.0  &   0.013&   0.419  &   $a$  \\
  103373.1&  C$_3$H      & $^2\Pi_{3/2}$ J=9/2--7/2 F=4--3 e &---  &   ---  &   ---    &   $a$   \\
   79151.1&  C$_3$N      & N=8--7 a                          &5.3  &   0.038&   0.935  & \\
   79170.0&  C$_3$N      & N=8--7 b                          &5.3  &   0.047&   1.084  &\\
   89045.7&  C$_3$N      & N=9--8 a                          &5.6  &   0.044&   1.454  &\\
   89064.4&  C$_3$N      & N=9--8 b                          &5.6  &   0.039&   1.212  &\\
   98940.0&  C$_3$N      & N=10--9 a                         &3.3  &   0.041&   1.170  &  \\
   98958.8&  C$_3$N      & N=10--9 b                         &3.3  &   0.037&   1.048  &\\
  108834.3&  C$_3$N      & N=11--10 a                        &3.2  &   0.043&   1.442  &\\
  108853.0&  C$_3$N      & N=11--10 b                        &3.2  &   0.044&   1.610  &\\
   76117.4&  C$_4$H      & N=8--7 a                          &4.6  &   0.037&   1.016  &   \\
   76157.0&  C$_4$H      & N=8--7 b                          &4.6  &   0.048&   1.276  &\\
   95150.3&  C$_4$H      & N=10--9 a                         &5.3  &   0.048&   1.385  &\\
   95188.9&  C$_4$H      & N=10--9 b                         &5.3  &   0.054&   1.358  &\\
  104666.6&  C$_4$H      & N=11--10 a                        &2.4  &   0.053&   1.507  &\\
  104705.1&  C$_4$H      & N=11--10 b                        &2.4  &   0.055&   1.438  &\\
   96478.3&  C$_4$H      & $\nu_7$=1 $^2\Pi_{3/2}$ J=21/2--19/2&4.8&   0.010&   0.289: &\\
  89188.5 &  HCO$^+$     & J=1--0                            &5.0  &   0.010&   0.267: &\\
  225697.8&  H$_2$CO     &3(1,2)-2(1,1)                      &6.0  &   0.023&   0.739  &\\
  244047.8&  H$_2$CS     &7(1,6)-6(1,5)                      &3.7  &   0.010&   0.239: &\\
   88315.1&  C$_5$H      & $^2\Pi_{1/2}$ J=37/2--35/2 e      &4.4  &   0.018&   0.543  &     $a$\\
   88320.9&  C$_5$H      & $^2\Pi_{1/2}$ J=37/2--35/2 f      &---  &   ---  &   ---    &     $a$\\
   88914.1&  C$_5$H      & $^2\Pi_{3/2}$ J=37/2--35/2 e      &5.5  &   0.015&   0.546  &     $a$\\
   88916.2&  C$_5$H      & $^2\Pi_{3/2}$ J=37/2--35/2 f      &---  &   ---  &   ---    &     $a$\\
   72783.8&  HC$_3$N     & J=8--7                            &4.8  &   0.158&   4.840  &        \\
   81881.5&  HC$_3$N     & J=9--8                            &4.8  &   0.198&   5.475  &\\
   90979.0&  HC$_3$N     & J=10--9                           &3.9  &   0.211&   6.860  &  \\
  100076.4&  HC$_3$N     & J=11--10                          &4.1  &   0.235&   7.335  &\\
  109173.6&  HC$_3$N     & J=12--11                          &3.6  &   0.284&   9.560  &    $c$\\
  227418.9&  HC$_3$N     & J=25--24                          &5.6  &   0.258&   8.069  &\\
  236512.8&  HC$_3$N     & J=26--25                          &5.0  &   0.248&   7.686  &\\
  245606.3&  HC$_3$N     & J=27--26                          &4.4  &   0.292&   9.496  &\\
  254699.5&  HC$_3$N     & J=28--27                          &6.5  &   0.274&   8.724  &\\
  263792.3&  HC$_3$N     & J=29--28                          &6.9  &   0.243&   7.705  &\\
  218860.6&  HC$_3$N     & $\nu_7$=1$^{1e} $J=24--23         &3.4  &   0.014&   0.457  &\\
  219173.6&  HC$_3$N     & $\nu_7$=1$^{1f} $J=24--23         &3.4  &   0.018&   0.498  &\\
  227977.1&  HC$_3$N     & $\nu_7$=1$^{1e} $J=25--24         &5.4  &   0.019&   0.243  &\\
  228303.0&  HC$_3$N     & $\nu_7$=1$^{1f} $J=25--24         &5.9  &   0.023&   0.524  &\\
  237093.2&  HC$_3$N     & $\nu_7$=1$^{1e} $J=26--25         &5.3  &   0.027&   0.264  &\\
  237432.0&  HC$_3$N     & $\nu_7$=1$^{1f} $J=26--25         &5.3  &   0.011&   0.220  &\\
  246208.9&  HC$_3$N     & $\nu_7$=1$^{1e} $J=27--26         &4.2  &   0.023&   0.423  &\\
  246560.7&  HC$_3$N     & $\nu_7$=1$^{1f} $J=27--26         &4.2  &   0.029&   0.434  &\\
  255324.3&  HC$_3$N     & $\nu_7$=1$^{1e} $J=28--27         &6.3  &   0.017&   0.209  &\\
  255689.1&  HC$_3$N     & $\nu_7$=1$^{1f} $J=28--27         &6.3  &   0.023&   0.210  &\\
  264439.3&  HC$_3$N     & $\nu_7$=1$^{1e} $J=29--28         &6.4  &   0.047&   0.473  &\\
  264817.0&  HC$_3$N     & $\nu_7$=1$^{1f} $J=29--28         &6.4  &   0.024&   0.314  &\\
   79350.5&  H$^{13}$CCCN& J=9--8                            &5.0  &   0.017&   0.465  &\\
   88166.8&  H$^{13}$CCCN& J=10--9                           &5.3  &   0.038&   0.292: &\\
   96983.0&  H$^{13}$CCCN& J=11--10                          &4.3  &   0.018&   0.393  & \\
  105799.1&  H$^{13}$CCCN& J=12--11                          &3.4  &   0.019&   0.396  &\\
  229203.1&  H$^{13}$CCCN& J=26--25                          &5.5  &   0.022&   0.200  &\\
   81534.1&  HC$^{13}$CCN& J=9--8                            &4.5  &   0.015&   0.542  & $d$ \\
   90593.1&  HC$^{13}$CCN& J=10--9                           &4.9  &   0.017&   0.794  & $d$ \\
   99651.8&  HC$^{13}$CCN& J=11--10                          &3.6  &   0.023&   0.758  & $d$ \\
  108710.5&  HC$^{13}$CCN& J=12--11                          &6.7  &   0.025&   1.531  &  $d$ \\
  253619.1&  HC$^{13}$CCN& J=28--27                          &6.4  &   0.022&   0.541  &  $d$ \\
   71889.6&  HC$_5$N     & J=27--26                         &9.0   &   0.042&   1.204  &\\
   74552.0&  HC$_5$N     & J=28--27                         &6.0   &   0.041&   1.402  &\\
   77214.4&  HC$_5$N     & J=29--28                         &7.8   &   0.047&   1.044  &\\
   79876.7&  HC$_5$N     & J=30--29                         &5.6   &   0.038&   0.645  &\\
   82539.0&  HC$_5$N     & J=31--30                         &4.6   &   0.036&   0.909  &\\
   85201.3&  HC$_5$N     & J=32--31                         &6.1   &   0.036&   0.774  &\\
   87863.6&  HC$_5$N     & J=33--32                         &10.4  &   0.047&   0.764: &\\
   90525.9&  HC$_5$N     & J=34--33                         &5.2   &   0.027&   0.894  &\\
   93188.1&  HC$_5$N     & J=35--34                         &4.4   &   0.026&   0.675  &\\
   95850.3&  HC$_5$N     & J=36--35                         &3.5   &   0.028&   0.765  &\\
   98512.5&  HC$_5$N     & J=37--36                         &2.9   &   0.023&   0.638  &\\
  101174.7&  HC$_5$N     & J=38--37                         &3.2   &   0.023&   0.738  &\\
  103836.8&  HC$_5$N     & J=39--38                         &1.9   &   0.026&   0.663  &\\
  106498.9&  HC$_5$N     & J=40--39                         &4.1   &   0.021&   0.800  &\\
  109161.0&  HC$_5$N     & J=41--40                         &---   &   ---  &   ---    &  $c$\\
   73588.8&  CH$_3$CN    &4(1)--3(1)                        &5.6   &  0.036 &  0.869   & $a$ \\
   73590.2&  CH$_3$CN    & 4(0)--3(0)                       &5.6   &   ---  &   ---    & $a$  \\
   91985.3&  CH$_3$CN    &5(1)--4(1)                        &4.0   &  0.028 &  1.186   & $a$ \\
   91987.1&  CH$_3$CN    & 5(0)--4(0)                       &---   &   ---  &   ---    & $a$  \\
  110381.4&  CH$_3$CN    &6(1)--5(1)                        &3.4   &  0.041 &  1.697   & $a$ \\
  110383.5&  CH$_3$CN    & 6(0)--5(0)                       &---   &   ---  &   ---    & $a$  \\
  220641.1&  CH$_3$CN    &12(5)--11(5)                      &7.5   &  0.018 &  0.418:  & \\
  220679.3&  CH$_3$CN    &12(4)--11(4)                      &7.5   &  0.018 &  0.156:  & \\
  220709.0&  CH$_3$CN    &12(3)--11(3)                      &7.5   &  0.029 &  0.637   & \\
  220730.3&  CH$_3$CN    &12(2)--11(2)                      &7.5   &  0.064 &  2.154   & $a$ \\
  220743.0&  CH$_3$CN    &12(1)--11(1)                      &  --- &  ---   &  ---     & $a$ \\
  220747.3&  CH$_3$CN    &12(0)--11(0)                      &  --- &  ---   &  ---     & $a$ \\
  239022.9&  CH$_3$CN    & 13(5)-12(5)                      &6.7  &   0.012&   0.127: &  \\
  239064.3&  CH$_3$CN    & 13(4)-12(4)                      &6.7  &   0.013&   0.195: &  \\
  239096.5&  CH$_3$CN    & 13(3)-12(3)                      &6.7  &   0.029&   0.392  &  \\
  239119.5&  CH$_3$CN    & 13(2)-12(2)                      &6.7  &   0.070&   3.010  &  $a$ \\
  239133.3&  CH$_3$CN    & 13(1)-12(1)                      &---   &   ---  &   ---    &  $a$ \\
  239137.9&  CH$_3$CN    & 13(0)-12(0)                      &---   &   ---  &   ---    &  $a$ \\
  257403.6&  CH$_3$CN    & 14(5)-13(5)                      &5.8   &   0.006&   0.030: &  \\
  257448.1&  CH$_3$CN    & 14(4)-13(4)                      &5.8  &   0.014&   0.086: &  \\
  257482.8&  CH$_3$CN    & 14(3)-13(3)                      &5.8  &   0.032&   0.419: &   \\
  257507.6&  CH$_3$CN    & 14(2)-13(2)                      &5.8  &   0.077&   2.400  & $a$ \\
  257522.4&  CH$_3$CN    & 14(1)-13(1)                      &---   &   ---  &   ---    & $a$  \\
  257527.4&  CH$_3$CN    & 14(0)-13(0)                      &---    &   ---  &   ---    & $a$  \\
  223938  &  U           &                                  &4.3  &   0.021&  0.541   &\\
  264191  &  U           &                                  &6.3  &   0.029&  0.540   & \\
\enddata
\begin{description}
\item $^{a}$ Unsolved hyperfine structure lines. 
\item $^{b}$ The  C$_2$H and SiC$_2$ lines are blended with each other. The former is stronger.
\item $^{c}$ The  HC$_3$H and HC$_5$H lines are blended with each other. The former is stronger.
\item $^{d}$ Blended with the corresponding transitions from HCC$^{13}$CN.
\end{description}
\end{deluxetable}

\begin{deluxetable}{lcccc@{\extracolsep{0.1in}}cc@{\extracolsep{0.1in}}cc}
\tablecaption{Excitation temperatures, column densities, and molecular abundances.
\label{col_crl2688}}
\tabletypesize{\scriptsize}
\tablewidth{0pt}
\tablehead{
\colhead{Species} & \colhead{$T_{\rm ex}$$^a$} & \multicolumn{3}{c}{$N$\,(cm$^{-2}$)} & \multicolumn{2}{c}{$f_X$}
& \multicolumn{2}{c}{$f_X/f_{\rm HC_3N}$}\\
\cline{3-5} \cline{6-7} \cline{8-9}
 & \,(K) & \colhead{This\,paper$^b$} & \colhead{F94$^c$} & \colhead{IRC+10216$^d$} & \colhead{This\,paper$^b$} & \colhead{IRC+10216$^e$} 
&\colhead{This\,paper$^b$} & \colhead{AFGL\,618$^f$} \\
}
\startdata
CN          &  \nodata     &$4.10\times10^{14}$&\nodata  &\nodata                   &$8.67\times10^{-6}$&$2.18\times10^{-6}$&21.49& 3.03   \\
$^{13}$CN   &  \nodata     &$2.38\times10^{13}$&\nodata  &\nodata                   &$2.85\times10^{-6}$&$7.60\times10^{-8}$&7.06 & 0.07   \\
CO        & \nodata     &$3.45\times10^{17} $&\nodata&\nodata                       &\nodata&\nodata	     &\nodata& 100   \\
$^{13}$CO & \nodata &$2.45\times10^{16} $&\nodata&\nodata                           &\nodata&\nodata	     &\nodata& 2.5    \\
C$^{17}$O & \nodata &$2.47\times10^{15} $&\nodata &\nodata                          &\nodata&\nodata	     &\nodata& 0.29        \\
C$^{18}$O & \nodata &$2.18\times10^{15} $&\nodata &\nodata                          &\nodata&\nodata	     &\nodata& 0.20        \\
CS        & \nodata     &$2.88\times10^{16} $&\nodata &\nodata                      &$7.04\times10^{-8}$&$1.10\times10^{-6}$&0.17 &  0.13   \\
$^{13}$CS & \nodata &$8.70\times10^{14} $&\nodata &\nodata                          &$1.69\times10^{-8}$&$2.20\times10^{-8}$&0.04 &  \nodata   \\ 
C$^{33}$S & \nodata &$2.63\times10^{14} $&\nodata &\nodata                          &$3.58\times10^{-9}$&\nodata	    &0.01 & \nodata   \\
C$^{34}$S & \nodata &$3.22\times10^{15} $&\nodata &\nodata                          &$5.40\times10^{-8}$&$3.70\times10^{-8}$&0.13 &  \nodata  \\
SiO         & 25.1         &$7.11\times10^{12}$&\nodata  &\nodata                   &$9.07\times10^{-9}$&$1.30\times10^{-7}$&0.02 &  0.01    \\
SiS         & 36.4         &$5.24\times10^{13}$&$4.5\times10^{13}$&\nodata          &$1.40\times10^{-7}$&$1.05\times10^{-6}$&0.35 &  \nodata   \\
C$_2$H      &  6.6         &$3.10\times10^{15}$&\nodata  & \nodata                  &$1.22\times10^{-6}$&$2.80\times10^{-6}$&3.03 &  0.50   \\
HCN       &\nodata     &$1.21\times10^{15} $&$4.32\times10^{15}$ & \nodata          &$6.17\times10^{-7}$&$1.40\times10^{-5}$&1.53 &  1.49  \\        
HC$^{15}$N& \nodata &$1.45\times10^{13} $&\nodata & \nodata                         &$2.91\times10^{-8}$&\nodata	    &0.07 &  \nodata     \\
H$^{13}$CN& \nodata &$1.52\times10^{14} $&\nodata & \nodata                         &$5.34\times10^{-7}$&$3.10\times10^{-7}$&1.32 &  0.03   \\
HNC         &  \nodata     &$5.03\times10^{13}$& $2.0\times10^{14}$ & \nodata       &$5.71\times10^{-7}$&$7.20\times10^{-8}$&1.42 &  0.17    \\
HN$^{13}$C  &  \nodata     &$1.60\times10^{12}$& \nodata &\nodata                   &$4.79\times10^{-9}$&$1.90\times10^{-9}$&0.01 &  0.004    \\
SiC$_2$   & 95.3    &$4.94\times10^{13} $ &\nodata  &2.9$\times10^{15}$             &$2.67\times10^{-7}$&$2.40\times10^{-7}$&0.66 &  \nodata      \\
C$_3$H    & 51.1    &$9.43\times10^{13} $ &\nodata  &$1.32\times10^{14}$            &$4.05\times10^{-7}$&$5.50\times10^{-8}$&1.00 &  0.02   \\
C$_3$N    & \nodata    &$1.08\times10^{14} $ &\nodata  &$4.54\times10^{14}$         &$5.36\times10^{-7}$&$3.00\times10^{-7}$&1.32 &  0.12   \\
C$_4$H    & 12.8    &$1.91\times10^{15} $ &$2.91\times10^{15}$  &$8.1\times10^{15}$ &$3.95\times10^{-6}$&$3.20\times10^{-7}$&9.79 &  1.0   \\
C$_5$H    & \nodata &$1.32\times10^{15} $ & \nodata&\nodata                         &$2.47\times10^{-7}$&\nodata            &0.61 &  0.005   \\
HCO$^+$   & \nodata &$1.67:\times10^{13}$ &\nodata &\nodata                        &$2.57:\times10^{-8}$&\nodata            &0.06 &  0.10  \\
H$_2$CO   & \nodata &$2.64\times10^{13} $ &\nodata &\nodata                         &$3.05\times10^{-8}$&\nodata            &0.08 &  0.13   \\
H$_2$CS   & \nodata &$1.63:\times10^{13}$ &\nodata &\nodata                        &$1.87:\times10^{-8}$&\nodata            &0.05 &  \nodata  \\
HC$_3$N   & 46.1    &$3.90\times10^{14} $ &$7.27\times10^{14}$ & $8.0\times10^{14}$ &$3.83\times10^{-7}$&$1.25\times10^{-6}$&1.0 &  1.0  \\
H$^{13}$CCCN& 32.4  &$2.37\times10^{13} $&\nodata & $2.0\times10^{13}$              &$3.20\times10^{-8}$&\nodata	    &0.08 &  \nodata  \\
HC$_5$N   & 27.8    &$2.06\times10^{14} $& $2.05\times10^{14}$  & \nodata           &$8.47\times10^{-8}$&$7.03\times10^{-6}$&0.21 & \nodata   \\
CH$_3$CN  & 28.0    &$5.85\times10^{13} $&\nodata&$2.6\times10^{13}$                &$7.60\times10^{-8}$&$6.60\times10^{-8}$&0.19 &  0.03 \\
\enddata
\tablenotetext{a}{A constant excitation temperature of 40\,K was assumed for the species for which the rotation diagrams cannot be obtainded.}
\tablenotetext{b}{For the species with optically thick emission, this gives the low limits.}
\tablenotetext{c}{From \citet{fukasaku94}.}
\tablenotetext{d}{From \citet{he08}.}
\tablenotetext{e}{From \citet{woods03}.}
\tablenotetext{f}{Those of the circumstellar shell in AFGL\,618; from \citet{pc07}.}
\end{deluxetable}

\begin{deluxetable}{llccc}
\tablecaption{Isotopic abundance ratios.
\label{isot}}
\tabletypesize{\scriptsize}
\tablewidth{0pt}
\tablehead{
\colhead{Isotopic ratio} & \multicolumn{2}{c}{AFGL\,2688} & \colhead{IRC+10216$^a$} & \colhead{Solar$^b$}\\
\cline{2-3}
& \colhead{Species}& \colhead{Value}\\
}
\startdata

$^{12}$C/$^{13}$C  & $^{12}$C$^{34}$S/$^{13}$C$^{32}$S & $61\pm30^c$ &   45$\pm3$       & 89   \\ 
                   &H$^{12}$C$_3$N/H$^{13}$CCCN      &$23\pm14$  & & \\
                   &$^{12}$CO/$^{13}$CO   & $>4.1$   &        &  \\ 
                   &$^{12}$CS/$^{13}$CS  & $>18.3$   &       & \\
                   &H$^{12}$CN/H$^{13}$CN& $>3.3$    &       & \\
$^{14}$N/$^{15}$N  & HC$^{14}$N/HC$^{15}$N  & $>24$  &  $>4400$       & 272 \\
                   & H$^{13}$C$^{14}$N/H$^{12}$C$^{15}$N & $<751^d$    &  & \\
$^{16}$O/$^{17}$O  & C$^{16}$O/C$^{17}$O  & $>37$    &  $840\pm200$       & 2681 \\
& $^{13}$C$^{16}$O/$^{12}$C$^{17}$O  & $<810^d$    &         & \\
$^{18}$O/$^{17}$O  & C$^{18}$O/C$^{17}$O  & $0.8\pm0.2$ &  $0.7\pm0.3$   &  5.37 \\
$^{32}$S/$^{34}$S  & C$^{32}$S/C$^{34}$S  & $>6.7$   & $21.8\pm2.6$   & 22.5 \\
$^{33}$S/$^{34}$S  & C$^{33}$S/C$^{34}$S  & $0.1:$   & $0.18\pm0.1$   & 0.18 \\
\enddata
\tablenotetext{a}{From the references in \citet{kah00}.}
\tablenotetext{b}{From \citet{lod03}.}
\tablenotetext{c}{Assume that the $^{34}$S/$^{32}$S ratio is solar.}
\tablenotetext{d}{Assume that the $^{12}$C/$^{13}$C ratio is smaller than the solar value.}
\end{deluxetable}

  \end{document}